\documentclass[twoside,twocolumn,9pt]{article}
\usepackage{extsizes}
\usepackage[super,sort&compress,comma]{natbib}
\usepackage[version=3]{mhchem}
\usepackage[left=1.5cm, right=1.5cm, top=1.785cm, bottom=2.0cm]{geometry}
\usepackage{balance}
\usepackage{times,mathptmx}
\usepackage{sectsty}
\usepackage{graphicx}
\usepackage{lastpage}
\usepackage[format=plain,justification=justified,singlelinecheck=false,font={stretch=1.125,small,sf},labelfont=bf,labelsep=space]{caption}
\usepackage{float}
\usepackage{fancyhdr}
\usepackage{fnpos}
\usepackage[english]{babel}
\addto{\captionsenglish}{%
  
}
\usepackage{array}
\usepackage{droidsans}
\usepackage{charter}
\usepackage[T1]{fontenc}
\usepackage[usenames,dvipsnames]{xcolor}
\usepackage{setspace}
\usepackage[compact]{titlesec}
\usepackage{hyperref}

\usepackage{subcaption}
\usepackage{multirow}
\usepackage{bm}
\usepackage{color}

\usepackage{epstopdf}

\definecolor{cream}{RGB}{222,217,201}

\begin{document}

\pagestyle{fancy}
\thispagestyle{plain}
\fancypagestyle{plain}{

\fancyhead[C]{\includegraphics[width=18.5cm]{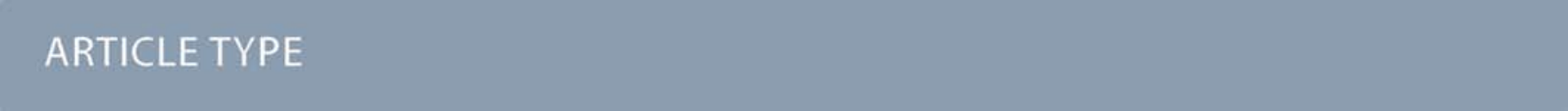}}
\fancyhead[L]{\hspace{0cm}\vspace{1.5cm}\includegraphics[height=30pt]{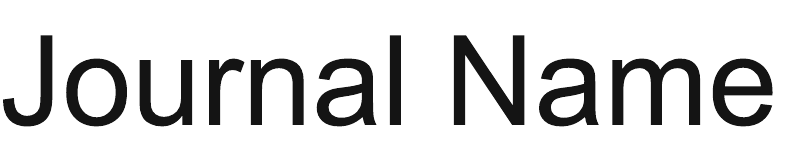}}
\fancyhead[R]{\hspace{0cm}\vspace{1.7cm}\includegraphics[height=55pt]{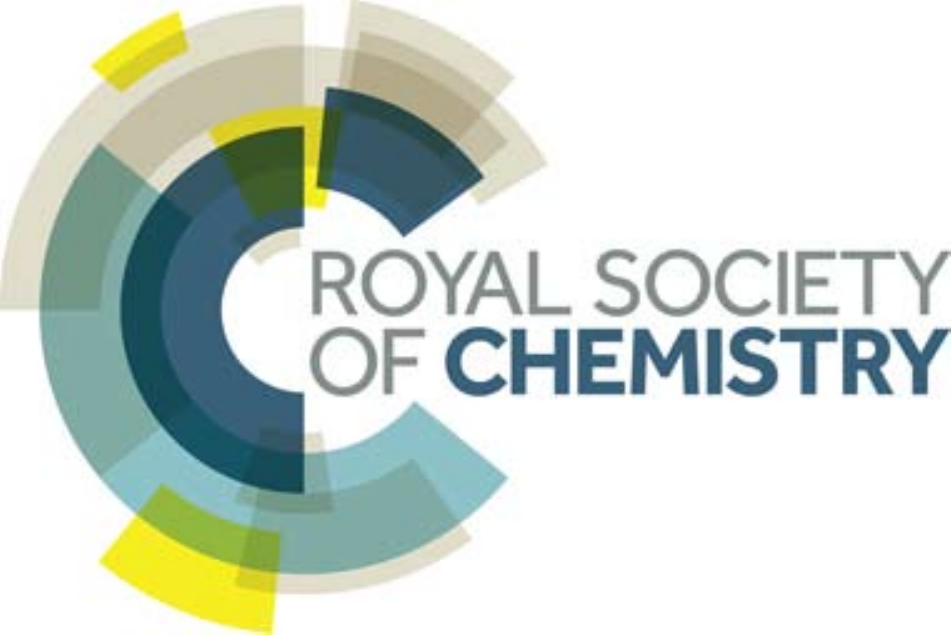}}
\renewcommand{\headrulewidth}{0pt}
}

\makeFNbottom
\makeatletter
\renewcommand\LARGE{\@setfontsize\LARGE{15pt}{17}}
\renewcommand\Large{\@setfontsize\Large{12pt}{14}}
\renewcommand\large{\@setfontsize\large{10pt}{12}}
\renewcommand\footnotesize{\@setfontsize\footnotesize{7pt}{10}}
\makeatother

\renewcommand{\thefootnote}{\fnsymbol{footnote}}
\renewcommand\footnoterule{\vspace*{1pt}%
\color{cream}\hrule width 3.5in height 0.4pt \color{black}\vspace*{5pt}}
\setcounter{secnumdepth}{5}

\makeatletter
\renewcommand\@biblabel[1]{#1}
\renewcommand\@makefntext[1]%
{\noindent\makebox[0pt][r]{\@thefnmark\,}#1}
\makeatother
\renewcommand{\figurename}{\small{Fig.}~}
\sectionfont{\sffamily\Large}
\subsectionfont{\normalsize}
\subsubsectionfont{\bf}
\setstretch{1.125} 
\setlength{\skip\footins}{0.8cm}
\setlength{\footnotesep}{0.25cm}
\setlength{\jot}{10pt}
\titlespacing*{\section}{0pt}{4pt}{4pt}
\titlespacing*{\subsection}{0pt}{15pt}{1pt}

\fancyfoot{}
\fancyfoot[LO,RE]{\vspace{-7.1pt}\includegraphics[height=9pt]{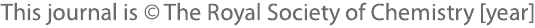}}
\fancyfoot[CO]{\vspace{-7.1pt}\hspace{13.2cm}\includegraphics{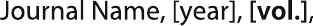}}
\fancyfoot[CE]{\vspace{-7.2pt}\hspace{-14.2cm}\includegraphics{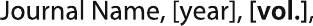}}
\fancyfoot[RO]{\footnotesize{\sffamily{1--\pageref{LastPage} ~\textbar  \hspace{2pt}\thepage}}}
\fancyfoot[LE]{\footnotesize{\sffamily{\thepage~\textbar\hspace{3.45cm} 1--\pageref{LastPage}}}}
\fancyhead{}
\renewcommand{\headrulewidth}{0pt}
\renewcommand{\footrulewidth}{0pt}
\setlength{\arrayrulewidth}{1pt}
\setlength{\columnsep}{6.5mm}
\setlength\bibsep{1pt}

\makeatletter
\newlength{\figrulesep}
\setlength{\figrulesep}{0.5\textfloatsep}

\newcommand{\topfigrule}{\vspace*{-1pt}%
\noindent{\color{cream}\rule[-\figrulesep]{\columnwidth}{1.5pt}} }

\newcommand{\botfigrule}{\vspace*{-2pt}%
\noindent{\color{cream}\rule[\figrulesep]{\columnwidth}{1.5pt}} }

\newcommand{\dblfigrule}{\vspace*{-1pt}%
\noindent{\color{cream}\rule[-\figrulesep]{\textwidth}{1.5pt}} }

\makeatother

\newcommand{\red}[1]{{\color{red} #1}}
\newcommand{\blue}[1]{{\color{blue} #1}}
\newcommand{\green}[1]{{\color{green} #1}}

\newcommand{\cm}{cm$^{-1}$}
\newcommand{\etal}{\textit{et al.}}

\newcommand{\onlinecite}[1]{\hspace{-1 ex} \nocite{#1}\citenum{#1}}

\twocolumn[
  \begin{@twocolumnfalse}
\vspace{3cm}
\sffamily
\begin{tabular}{m{4.5cm} p{13.5cm} }

\includegraphics{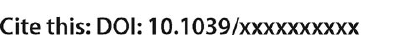} &
\noindent\LARGE{\textbf{A variationally computed room temperature line list for AsH$_{3}$ $^\dag$}} \\
\vspace{0.3cm} & \vspace{0.3cm}

\noindent\large{Phillip A. Coles,\textit{$^{a}$} Sergei N. Yurchenko,\textit{$^{a}$} Richard P. Kovacich,\textit{$^{b}$} James Hobby,\textit{$^{b}$} and Jonathan Tennyson,$^{\ast}$\textit{$^{a}$}}
 \\

\includegraphics{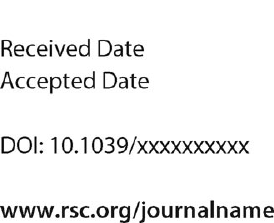} & \noindent\normalsize{Calculations are reported on the rotation-vibration energy levels of the arsine molecule with
associated transition intensities. A potential energy surface (PES) obtained
from \textit{ab initio} electronic structure calculations is refined to
experimental data, and the resulting energy levels display sub-wavenumber accuracy for all reliably known $J=0$ term values under 6500 cm$^{-1}$. After a small empirical adjustment of the band centres, our calculated ($J=1-6$) rovibrational states reproduce 578 experimentally derived energies with a root-mean-square error of  0.122 cm$^{-1}$.
Absolute line intensities are computed using the refined PES and a new
dipole moment surface (DMS) for transitions between states with energies up to 10~500 cm$^{-1}$ and rotational quantum number $J=30$. The computed DMS reproduces experimental line
intensities to within 10\% uncertainty for the $\nu_1$ and $\nu_3$ bands. Furthermore, our calculated absorption cross-sections display good agreement with the main absorption features recorded in Pacific Northwest National Laboratory (PNNL) for the complete range of $600-6500$ cm$^{-1}$. } \\

\end{tabular}

 \end{@twocolumnfalse} \vspace{0.6cm}

  ]

\renewcommand*\rmdefault{bch}\normalfont\upshape
\rmfamily
\section*{}
\vspace{-1cm}


\footnotetext{\textit{$^{a}$~~Department of Physics \& Astronomy, University College London, London WC1E~6BT, UK }}
\footnotetext{\textit{$^{b}$~~Servomex Ltd.,  Millbrook Industrial Estate, Crowborough TN6 3FB, UK }}

\footnotetext{\dag~Electronic Supplementary Information (ESI) available: including the final line list, partition function and other spectroscopic data
is available from the ExoMol website, www.exomol.com.
See DOI: 10.1039/b000000x/}




\section{Introduction}

Arsine (AsH$_3$) is a highly poisonous gas\cite{06Paxxxx.AsH3} which is the direct analogue molecular structure of ammmonia (NH$_3$) and phospine (PH$_3$).
Like these two gases it has been detected the atmospheres of the gas giant planets
Jupiter \cite{89NoGeKn.AsH3,90NoLaGe.AsH3} and Saturn \cite{89BeDrKe.AsH3}. It may therefore be expected
to be also present in the atmospheres of gas giant exoplanets.

Arsene is also important for industrial applications as high purity
arsine is widely used in the semiconductor manufacturing industry, for
example, in processing GaAs surfaces
\cite{05KhHaKu.AsH3,09ChHeAn.AsH3,15KoBaLe.AsH3}.  Given its highly
poisonous nature, with an exposure limit value of 50 ppb
mole-concentration\cite{EH40.AsH3}, the detection of AsH$_3$ escape at
such levels is an important safety requirement in this
industry\cite{06ChHsAg.AsH3}. It is also monitored in the polymer
industry as trace level arsine impurity in ethylene and propylene
monomer feedstock gases may contaminate the catalysts, resulting in
reduced quality and yield of the polymer products
\cite{05FeWaxx.AsH3}.

Arsine is also a trace atmospheric pollutant due
to emissions from various industrial processes, such as power
generation and smelting \cite{00MaMoxx.AsH3,00Maxxxx.AsH3}.  Routine
methods for arsine measurement in industry include gas chromatography,
electrochemical sensors, colorimetric sensors, and Fourier-transform
infrared spectroscopy. However, the development of high resolution
laser spectroscopy based measurements is a growing
area\cite{93StTrxx.AsH3,10Coxxxx.AsH3} for which detailed line lists
are required to model the high resolution absorption spectra.

While there have been a number of studies of the infrared and microwave spectrum of arsine,
there is no comprehensive line list for the system and there is a lack of information on the intensity
of many bands. The situation for absolute line intensities is particularly dire, with existing data confined solely to the measurements reported by Dana \etal~\cite{93DaMaTa.AsH3}. Previous attempts to model the global vibrational structure \cite{08SaLeUl.AsH3,03PlLeMo.AsH3}, and rovibrational sub-structures\cite{93UkChSh.AsH3,93UkMaWi.AsH3,97YaWaZh.AsH3,
98YaLiWa.AsH3,98WaLiWa.AsH3,98LiUlYu.AsH3,00HaUlOl.AsH3}, have focussed predominantly on Effective-Hamiltonians, which have limited predictive capability outside the fitted data. In addition, an \textit{ab initio} potential energy surface (PES) for AsH$_3$ was reported in 1995 by Breiding and Thiel in the form of the cubic anharmonic force field \cite{95BrThxx.AsH3} using relativistic effective core potentials (ECPs).

Considering the unsuitability of the current state of AsH$_3$ data for either exoplanet modelling, which necessitates completeness, or industrial monitoring, which necessitates accuracy, we decided to construct a comprehensive line list for arsine which
could be used for the applications mentioned above.
Our approach to constructing linelists, as exploited in the ExoMol project \cite{jt528,jt511},
uses potential energy surfaces which have been refined using spectroscopic data but
{\it ab initio} dipole moment surfaces, which have been shown to give highly accurate
predicted transition intensives \cite{jt613,jt721,jt744}. The approach has already been
used to compute line lists for the systems
NH$_3$ \cite{jt466,jt500,jt743}, PH$_3$ \cite{jt556,jt592}, SbH$_3$ \cite{10YuCaYa.SbH3}, SO$_3$ \cite{jt580} and  and a linelist for PF$_3$ is on the way \cite{19MaChYa.PF3}. These
studies were based on either a D$_{\rm 3h}$ symmetry \cite{jt466,jt500,jt743,jt658},
in which the tunneling of the so-called umbrella motion was explicitly considered, or C$_{\rm 3v}$ symmetry
\cite{jt556,jt592} in which the tunneling is neglected. Given the high barrier expected for AsH$_3$,
the reduced symmetry C$_{\rm 3v}$ approach is employed here.

\section{Potential energy surface}\label{PES}
\subsection{Electronic structure calculations}\label{elec_struc}
Accurate modelling of heavy elements in quantum chemistry is made particularly
challenging by increased relativistic effects, core-core electron correlation
and core-valence electron correlation compared to lighter elements.
Recovery of the correlation energy has benefited from the recent development of
explicitly correlated methods (F12/R12) \cite{07AdKnGe.method,09KnAdWe}, which
rapidly converge electronic energies towards the complete basis set (CBS) limit with increasing cardinal
number $n$, but must be used in conjunction with suitably optimised basis sets
for full effectiveness.
To account for scalar relativistic effects, effective core potentials (ECPs) are
a computationally inexpensive option. More rigorous treatment is possible with
the
Douglas-Kroll-Hess (DKH) Hamiltonian, however, so far no satisfactory way has
been found of incorporating the DKH Hamiltonian into the F12/R12 framework
\cite{10BiVaKl}. Although ECPs face the same theoretical hurdles, namely
non-commutability with the F12 correlation function, alternative treatments have
been found to work well \cite{08BiHoKl,11WeKnMa}. For heavy elements where both
F12-pseudopotential  and standard all-electron DKH based approaches are
possible, such as arsenic, the benefits of F12 must therefore be weighed
against the penalty of introducing an additional scalar-relativistic
approximation. Peterson \cite{11PeKrSt} showed that complete basis set (CBS)
extrapolated CCSD(T)/aug-cc-pwCVnZ-PP
pseudopotential calculations performed almost identically to their DKH
all-electron counterparts in a series of molecular benchmark calculations for
post-3d main group elements, including the As$_2$, AsF, AsCl and AsN molecules.
They go on to develop a new family of F12-specific cc-pVnZ-PP-F12 basis sets to
be used at the CCSD(T)-F12 level, which yield accuracy comparable to the 2-3
times larger aug-cc-pwCV(n+2)Z-PP basis sets used at standard CCSD(T) level
\cite{11PeKrSt,14HiPexx}.
Their pseudopotential-F12 optimised approach is the one followed in this work.

All electronic structure calculations were performed using MOLPRO
\cite{12WeKnKn.methods} and employed the explicitly correlated coupled
cluster method CCSD(T)-F12b \cite{07AdKnGe.method,09KnAdWe} with implicit treatment
of scalar-relativistic effects via replacement of 10 core electrons with a
pseudopotential (PP). Calculations were carried out in the frozen core
approximation and utilized the correlation consistent quadruple-zeta, PP-F12
optimised basis set of Hill \etal~\cite{14HiPexx} (cc-pVQZ-PP-F12) to represent
the As electronic wavefunction, and cc-pVQZ-F12 basis sets for the H atoms. Density fitting (DF) for the
2-electron (MP2FIT) and exchange term (JKFIT) integrals employed the
cc-pVTZ-PP-F12/MP2Fit and def2-QZVPP/JKFIT basis sets respectively, and for the
resolution of the identity of the many-electron F12 integrals (OPTRI) we used
the VTZ-PP-F12/OPTRI basis set. For the geminal exponent $\gamma$, a value of
1.4 a$_0^{-1}$ was used as recommended by Hill \etal.
All calculations were performed on the ground electronic state, which is sufficiently uncoupled from higher
electronic excitations that both adiabatic and non-adiabatic effects are expected to be very small \cite{12AlBuLi.AsH3}.

\begin{table}[h!]
\centering
\caption{Equilibrium energies calculated at the CCSD(T) level of theory using different basis sets and Hamiltonians.}
  \begin{tabular}{ l  r }
    \hline
Basis   &   Energy/$E_{\rm h}$  \\ \hline
AVQZ    & -2236.17795527 \\
AVQZ-DK & -2261.02359376 \\
AVQZ-PP &  -333.14700414 \\
AV5Z    & -2236.18098311\\
AV5Z-DK & -2261.02786843\\
AV5Z-PP &  -333.14983002\\
    \hline
  \end{tabular}
\label{table:Emin}
\end{table}

\begin{figure*}[!h]
\centering
\begin{subfigure}{.5\textwidth}
  \centering
  \includegraphics[width=.9\linewidth]{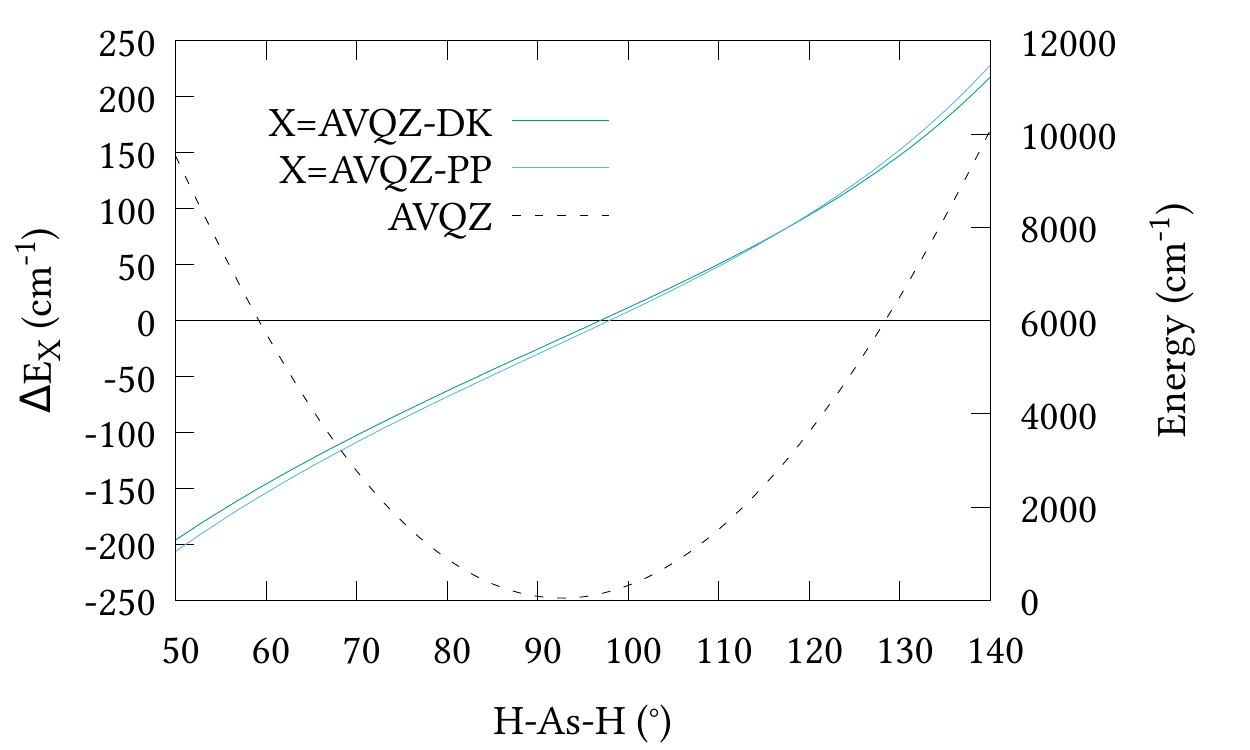}
  \label{fig:sub1}
\end{subfigure}%
\begin{subfigure}{.5\textwidth}
  \centering
  \includegraphics[width=.9\linewidth]{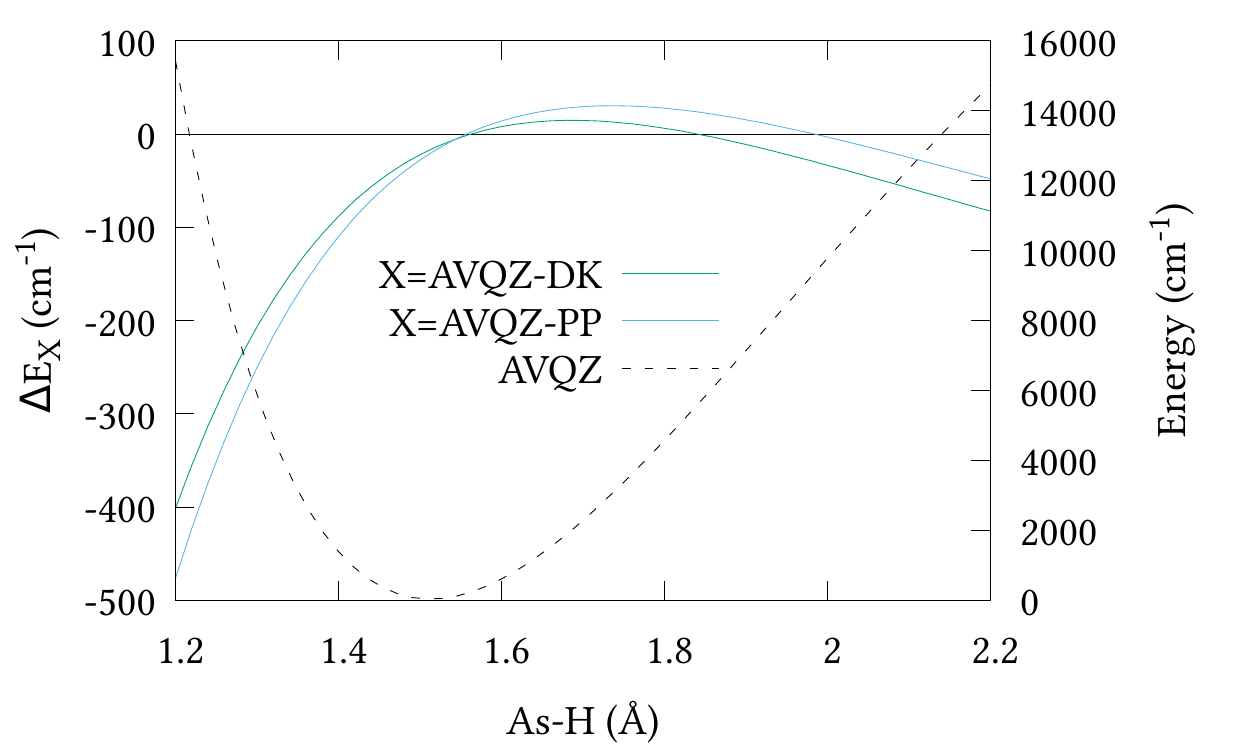}
  \label{fig:sub2}
\end{subfigure}
\caption{One dimensional cuts of the relativistic corrections for the ($r_1=r_2=r_3=1.51$ \AA; $\alpha_1=\alpha_2=92.1^{\circ}$; $50 \leq \alpha_3 \leq 140^{\circ}$) bond angle and ($r_1=r_2=1.51$ \AA;$1.2\leq r_3\leq 2.2$ \AA ; $\alpha_1=\alpha_2=\alpha_2=92.1^{\circ}$) bond length displacements.}
\label{fig:rel_corr}
\end{figure*}

To initially qualify the importance of including scalar relativistic effects in
our calculations, relativistic corrections $\Delta E_X$ along 1-dimensional cuts
of the potential energy surface PES were calculated (shown in
Fig.~\ref{fig:rel_corr}, along with cuts through the AVQZ surface for reference).
This was done by
by first shifting the potential energy curves by their respective energies at
equilibrium, listed in table~\ref{table:Emin}, where we note the relatively small absolute energies of the ECP based calculations owing to their implicit treatment of 10 core electrons. We then have $\Delta E_X =E_{X_{1}}-E_{X_{2}}$ where
$E_{X_{1}}=$AV$n$Z-PP/AV$n$Z-DK and $E_{X_{2}}=$AV$n$Z for the
pseudopotential/all-electron calculations. Here,  and in all subsequent DKH calculations, the DKH Hamiltonian has been expanded to 8$^{\rm th}$--order (DKH8) using optimal unitary parametrisation. In Fig.~\ref{fig:rel_corr} only quadruple-zeta ($n=4$) results
are presented as they were seen to differ by no more than 3 cm$^{-1}$ from the
respective 5-zeta ($n=5$) curves. Clearly the inclusion of scalar relativistic
effects are important, and both approximations have a similar effect on the
total energy. However, the pseudopotential approximation tends to raise the
energy at stretched geometries and lower the energy at contracted geometries,
relative to the all-electron calculations.
It is difficult to asses the effect of this difference within the Hill
\etal~\cite{14HiPexx} regime for a full-dimensional PES.
We therefore opted to generate a second 6D surface at the
AVQZ-DKH8 level of theory (henceforth denoted AVQZ-DK), to provide a benchmark for our VQZ-PP-F12 based \textit{ab initio}
nuclear motion calculations. The results of these are presented in section~\ref{nucl_motion}.

It is known that beyond 4$^{\rm th}$--order expansion the DKH
Hamiltonian depends slightly on the chosen paramterization of the
unitary transformations applied\cite{11NaHixx}. Thus, additional DKH4
calculations were performed on a subset of the DKH8 PES geometries.
Between 4$^{\rm th}$ and 8$^{\rm th }$--order the resulting electronic
energies were seen to differ by less than 0.1 cm$^{-1}$ above their
equilibrium values. For the purpose of benchmarking the ECP based
nuclear motion calculations, therefore, this dependency is not expected
to be significant.

\subsection{Nuclear geometry grid}\label{geom_grid}
Our grid of nuclear geometries was built by combining 1D$-$6D sub-grids. Our 1D
grid consisted of a cut along the $r_1=r_2=r_3$ stretch with
$\alpha_1=\alpha_2=\alpha_3=\alpha_{eq}$, and a cut along the
$\alpha_1=\alpha_2=\alpha_3$ bend with $r_1=r_2=r_3=r_{\rm eq}$. Each additional
degree of freedom was then added by allowing either an additional As-H bond
length or H-As-H bond angle to vary. Because this method causes the number of
points to grow so rapidly, the range and intervals of $r$ and $\alpha$ were
reduced with each increasing degree of freedom using the 1D cuts as a guide.
This also helped to limit the range of electronic energies generated, as large
distortions in geometry can lead to unnecessarily high values of energy that
are not needed in the fit.

In order to ensure each grid point was fully unique we applied the C$_{\rm 3v}$
molecular symmetry group transformations prior to computing the electronic
energy. If two grid points were transformed into one another, then one was
discarded. Finally, any energetically sparse regions were filled by generating
additional geometries that were estimated to fall within our desired range. The
1D cuts provided an initial guide to the electronic energy, then intermediate
versions of our PES were used to more accurately choose geometries. Our final
grid consisted of 39~873 nuclear geometries within the range $1.10\leq r_i\leq 3.74$ \AA\
and $37^\circ\leq \alpha_i\leq 130^\circ$, with electronic energies extending to 27~000
cm$^{-1}$, although $\sim$ 38~000 of these were below 10,000 cm$^{-1}$ ($1.25\leq r_i\leq 1.9$ \AA\
and $60^\circ\leq \alpha_i\leq 126^\circ$). The additional
points in the 10~000$-$27~000 cm$^{-1}$ range predominantly belonged to the As-H
dissociative stretch, which is where holes commonly appear if the function is
not suitably constrained at high energy. Grid points for our AVQZ-DK reference
PES were chosen by randomly sampling 16~396 equally energetically distributed
points from our VQZ-PP-F12 grid, which spanned the bond lengths $1.3\leq r_i\leq 1.8$ \AA, bond angles
$65^\circ\leq \alpha_i\leq 130^\circ$ and energies below 15~000 cm$^{-1}$.

Each grid point computed at the CCSD(T)-F12b/cc-pVQZ-PP-F12 level took
approximately 10-15 minutes to compute on UCL's Legion computer cluster. This was
increased to 20-30 mins for the DKH Hamiltonian-based calculations, owing to the
increased computational demand of explicitly treating the 10 core electrons.

\subsection{Analytic representation}\label{PES_rep}

To represent the PES analytically we used the same functional form as for NH$_3$
and PH$_3$ \cite{jt503}. The potential is represented as a polynomial expansion
\begin{equation} \label{eq:Vfunc}
\begin{aligned}
V(\xi_1,\xi_2,\xi_3, &\xi_{4a},\xi_{4b};\sin(\bar{\rho}))= V_e+V_0\sin(\bar{\rho})+\sum_{i}F_i\sin(\bar{\rho})\xi_i \\
 & + \sum_{i\le j}F_{ij}\sin(\bar{\rho})\xi_i\xi_j + ...\\
 & + \sum_{i\le j\le k\le l\le m\le n}F_{ijklmn}\sin(\bar{\rho})\xi_i\xi_j\xi_k\xi_l\xi_m\xi_n
\end{aligned}
\end{equation}
in terms of the internal coordinates
\begin{eqnarray}
\label{eq:icoord1}
\xi_k&=&1-\exp(-a(r_k-r_{\rm eq})), \quad (k=1,2,3), \\
\label{eq:icoord2}
\xi_{4}&=&(2\alpha_1-\alpha_2-\alpha_3)/\sqrt{6}, \\
\xi_{5}&=&(\alpha_2-\alpha_3)/\sqrt{2}, \\
\label{eq:sin:brho}
   \sin {\bar \rho} \, &=& \, \frac{2}{\sqrt{3}} \,\sin [
     (\alpha_{1} + \alpha_{2} + \alpha_{3}) /6] \hbox{.}
\end{eqnarray}
In Eq.~(\ref{eq:Vfunc})
\begin{equation}\label{eq:fparams}
F_{ij...}\sin(\bar{\rho})=\sum_{s=0}^Nf_{ij...}^{(s)}(\sin(\rho_{\rm eq})-\sin(\bar{
\rho}))^s
\end{equation}
and $r_k$ is the As-H$_k$ bond length, $\alpha_j$ is the $j^{th}$ H-As-H bond
angle (opposite to the $j$th bond), $r_{\rm eq}$ is the equilibrium value of $r_k$, $a$ is a molecular parameter, and $\rho_{\rm eq}$ is the equilibrium value of the umbrella mode $\bar\rho$. $V_0$ represents the pure
inversion potential and $f_{ij...}^{(s)}$ are the fitting parameters.
Points were given energy ($E_i$) dependant weights ($w_i$)

\begin{equation} \label{eq:wghts}
w_i=\dfrac{2}{1+e^{2\times 10^{-4}\times E_i}}
\end{equation}
as used by Polyansky~\etal~\cite{jt634}.  We could usefully fit terms
in the potential up to 5$^{th}$ order resulting in a root-mean-square
(RMS) deviation of 0.7 cm$^{-1}$ for the 39~678 nuclear geometries.
For our all-electron reference PES, the weighted RMS error increased
to 1.2 cm$^{-1}$, most likely due to the proportionally fewer points
close to equilibrium. However it should be noted that adding more
points to the fit had little effect on the computed vibrational term
values reported in section~\ref{nucl_motion}.

The final set of expansion parameters for our \textit{ab initio} PES is included
in the supplementary material, along with our grid of nuclear geometries and an
Fortran 90 routine to evaluate the analytic expression.

\section{Nuclear motion calculations}\label{nucl_motion}

\begin{table*}[!h]
\caption{Differences between experimentally derived band centres and our calculated values computed using all-electron
DKH and pseudopotential-F12 based PESs. All numerical values are energies given in units of cm$^{-1}$.}
  \begin{tabular}{ l  l l c c c c c}
    \hline
    Band          & Symmetry   & Obs.      & VQZ-PP-F12 & AVQZ-DK & Obs$-$Calc$_{\rm PP-F12}$ & Obs$-$Calc$_{\rm DK}$\\
 \hline
    $\nu_2$             & A$_1$ &  906.752 & 904.812	& 905.058	&	1.940	&	1.694	\\
    $\nu_4$             & E     &  999.225 & 994.460	& 994.132	&	4.765	&	5.093	\\
    $2\nu_2$            & A$_1$ & 1806.149 & 1802.443   & 1802.451	&	3.706   &	3.698	\\
    $\nu_2+\nu_4$       & E     & 1904.115 & 1897.551	& 1897.465  &	6.564	&	6.650	\\
    $2\nu_4^{l=0}$      & A$_1$ & 1990.998 & 1982.116	& 1981.574	&	8.882	&	9.424	\\
    $2\nu_4^{l=2}$      & E     & 2003.483 & 1988.246   & 1987.651  &   15.237  &   15.832  \\
    $\nu_1$             & A$_1$ & 2115.164 & 2108.659   & 2105.000	&	6.505   &	10.164	\\
    $\nu_3$             & E     & 2126.432 & 2116.469   & 2112.542	&	9.963	&	13.890	\\
    $\nu_1+\nu_2$       & A$_1$ & 3013$^a$ & 3006.718   & 3002.875  &   6.3     &   10.1    \\
    $\nu_1+\nu_4$       & E     & 3102$^a$ & 3089.255   & 3084.866  &   12.7    &   17.1    \\
    $2\nu_1$            & A$_1$ & 4166.772 & 4151.833	& 4143.187	&	14.939	&	23.585	\\
    $\nu_1+\nu_3$       & E     & 4167.935 & 4152.229	& 4143.527	&	15.706	&	24.408	\\
    $2\nu_3^{l=0}$      & A$_1$ & 4237.700 & 4222.006	& 4214.312	&	15.694	&	23.388	\\
    $2\nu_3^{l=2}$      & E 	& 4247.720 & 4229.805	& 4221.816	&	17.915	& 	25.904	\\
    $2\nu_1+\nu_2$      & A$_1$ & 5057$^a$ & 5041.541   & 5030.916  &   15.5    &   26.1    \\
    $\nu_1+\nu_2+\nu_3$ & E     & 5057$^a$ & 5041.191   & 5030.920  &   15.8    &   26.1    \\
    $2\nu_1+\nu_4$      & E     & 5128$^a$ & 5111.286   & 5100.477  &   16.7    &   27.5    \\
    $2\nu_3^0+\nu_2$    & A$_1$ & 5128$^a$ & 5113.615   & 5104.249  &   14.4    &   23.8    \\
    $\nu_1+\nu_3+\nu_4$ & E     & 5158$^a$ & 5137.282   & 5127.176  &   20.7    &   30.8    \\
    $\nu_1+\nu_3+\nu_4$ & A$_1$ & 5158$^a$ & 5137.555   & 5127.471  &   20.4    &   30.6    \\
    $3\nu_1$            & A$_1$ & 6136.340 & 6116.822   & 6101.231  &   19.518  &   35.109  \\
    $2\nu_1+\nu_3$      & E     & 6136.330 & 6116.793   & 6101.192  &   19.537  &   35.138  \\
    $\nu_1+2\nu_3^{l=0}$& A$_1$ & 6275.830 & 6257.116   & 6243.540  &   18.714  &   32.290  \\
    $\nu_1+2\nu_3^{l=2}$& E     & 6282.350 & 6261.282   & 6247.600  &   21.068  &   34.750  \\
    $3\nu_3^{l=1}$      & E     & 6294.710 & 6270.037   & 6256.059  &   24.673  &   38.651  \\
    $3\nu_3^{l=3}$      & A$_1$ & 6365.950 & 6340.980   & 6327.902  &   24.970  &   38.048  \\
    \hline
  \end{tabular}
\mbox{}\\
$^a$ experimental uncertainties of Halonen \etal~\cite{92HaHaBu.AsH3} are estimated to be 2 cm$^{-1}$ or more.
\label{table:p10j0}
\end{table*}

To calculate rovibrational energy levels we used the variational
nuclear motion program TROVE. The general methodology of TROVE is well
documented \cite{TROVE,17YuYaOv.methods}, with its specific
application to XY$_3$-type molecules detailed in
Ref.~\onlinecite{jt466}, and so only a brief description is
provided here.

Rovibrational basis functions are constructed as symmetrised linear
combinations of 1D primitive-basis-function products
\begin{equation}\label{rv_basis}
|\nu,J,K,m,\tau_{\rm rot}\rangle=[|J,K,m,\tau_{\rm rot}\rangle|n_1\rangle|n_2\rangle|n_3\rangle|n_4\rangle|n_5\rangle|n_6\rangle]^{\Gamma_{\rm ir}}
\end{equation}
where the 1D stretching functions
($|n_1\rangle$,$|n_2\rangle$,$|n_3\rangle$) and bending functions
($|n_4\rangle$,$|n_5\rangle$,$|n_6\rangle$) are obtained by solving
the corresponding one-dimensional Schr{\"o}dinger equations using the
Numerov-Cooley approach\cite{24Numerov.method,61Cooley.method} for the
stretches, and 1D harmonic oscillator eigenfunctions for the bends. The
rotational basis set is built from rigid-rotor functions. In the above
equation, $\Gamma_{\rm ir}$ represents one of the irreducible
representations of C$_{\rm 3v}$ spanned by $|\nu,J,K,m,\tau_{\rm
  rot}\rangle$. For details of the symmetrisation procedure in TROVE
the reader is directed to Ref.~\cite{17YuYaOv.methods}. A multi-step
contraction scheme was employed to limit the vibrational, then
rovibrational basis set size. This is outlined in the following
paragraphs and in section \ref{rovib}.

Owing to the structural similarities between AsH$_3$ and other
XY$_3$-type molecules which have been investigated in the past,
variational calculations could be performed with relative ease once a
PES and DMS had been constructed, and required only a few molecule
specific parameters to be defined: atomic mass, molecular symmetry
group, Z-matrix, equilibrium parameters, and the definition of our 1-D
grids upon which the wavefunctions are evaluated. In fact, this is
true for any 3-5 atom molecule, providing the molecular symmetry (MS)
group has been programmed into TROVE, this includes a recent extension
to treat linear molecules \cite{jt730}. Currently TROVE allows for the
following MS groups: C$_{\rm 2v}$, C$_{\rm 3v}$, D$_{\rm 3h}$, D$_{\rm
  2d}$, C$_{\rm 2h}$, T$_{\rm d}$, D$_{\rm 2h}$, as well as D$_{n{\rm
    h}}$ with $n>1$ \cite{18ChJeYu} (see also
Ref.~\onlinecite{17YuYaOv.methods}, where the TROVE treatment of the
symmetry groups is described). Here we use C$_{\rm 3v}$ symmetry which
means that the inversion mode is not fully represented and any
tunneling splitting is neglected. We note that study of deuterated arsine,
AsH$_2$D could also be conducted using the same Born-Oppenheimer PES but
would require a different symmetrisation procedure in TROVE. Such
a study is possible but the lower
symmetry and denser vibrational spectrum associated with the D atom would
make such a calculation computationally more expensive.

As arsenic has only one stable isotope, $^{75}$As, all reported nuclear motion calculations were performed for $^{75}$AsH$_3$.

Within the limitations of our PES, the accuracy of our variational
calculation is determined predominantly by i) the size of our
nuclear-motion basis set; and ii) our Taylor-type expansion of the
kinetic energy operator $\hat{T}$ and our re-expansion of the
potential function $V$ in terms of linearised coordinates.  The
former, we choose to restrict via a polyad number P. The polyad number
is an integer that represents the total quanta of vibrational
excitations in terms of the lowest energy fundamental. For AsH$_3$
\begin{equation}
\mathrm{P}=2(n_1+n_2+n_3)+n_4+n_5+n_6
\end{equation}
where $n_1+n_2+n_3$ is the total number of stretching quanta and
$n_4+n_5+n_6$ is the total number of bending quanta, corresponding to
the primitive functions $|n_i\rangle$ ($i=1,...,6$) in
Eq.(\ref{rv_basis}). For our comparison of the VQZ-PP-F12 and AVQZ-DK
based \textit{ab initio} PESs we chose to include in our variational
calculations all vibrational states with P $\le$ P$_\mathrm{max}=$14
as used previously for NH$_3$ and PH$_3$ \cite{jt466,jt556}. This
resulted in our vibrational eigenvalues converged to within 0.1
cm$^{-1}$ below 6000 cm$^{-1}$ for the stretches, and as much as 3
cm$^{-1}$ for the bending overtones. Our $\hat{T}$ and $V$ expansions
we take to 6$^{\rm th}$ and 8$^{\rm th}$ order respectively.
Increasing these to 8$^{\rm th}$ and 10$^{\rm th}$ order changed the
vibrational term values reported throughout this work by $< 0.1$ \cm\
for the stretches, and $< 0.7$ \cm\ for the bends. For highly excited
bending overtones, such as the 5 and 6--quanta bends, the convergence
error due to our $\hat{T}$ and $V$ expansions may be several
wavenumbers.

Table \ref{table:p10j0} shows the 26 lowest-lying experimentally
derived vibrational states compared to our calculations. Term values
known to sub-wavenumber accuracy are taken from Sanzharov \textit{et
  al}~\cite{08SaLeUl.AsH3}; the remaining eight bands are from the
work by Halonen \etal~\cite{92HaHaBu.AsH3} and have an estimated 2
cm$^{-1}$ uncertainty, although this may be larger for the 5050$-$5200
cm$^{-1}$ bands \cite{95LuKaHa.AsH3}. Using the VQZ-PP-F12 and AVQZ-DK
PESs the experimentally derived values of the four fundamentals are
reproduced to within 10 cm$^{-1}$ and 14 cm$^{-1}$ respectively.
Whilst far from the accuracy achieved in previous studies of NH$_3$
and PH$_3$, our results are comparable to the achievements of Nikitin
\etal\ in their recent {\it ab initio} study of GeH$_4$
\cite{16NiReRo.GeH4}, and we deem it reasonable considering the
greater contribution of relativistic effects, core-core electron
correlation and core-valence electron correlation associated with
heavier atoms. For the overtones and combination bands the quality of
our \textit{ab initio} predictions steadily decreases in proportion to
the error on the fundamentals, except for the $2\nu_4^{l=0}$ band
which is independently examined in section \ref{rovib}. Most
importantly, the VQZ-PP-F12 surface consistently and significantly
outperforms the VQZ-DK surface. Given the factor of 2 reduction in
computational time, this highlights the value of the work by Hill,
Peterson and co-authors~\cite{14HiPexx,11PeKrSt}.

\section{Refinement}\label{refinement}
In order to achieve so-called `spectroscopic' accuracy in our variational nuclear motion calculations it is common practice to empirically refine the chosen \textit{ab initio} PES to experimental data. In this case our chosen starting point PES is the VQZ-PP-F12 surface. A theoretical description of the refinement procedure has been previously
reported for the case of NH$_3$ \cite{jt503}, and our method for AsH$_3$
is similar. Namely, small corrections to the parameters $f_{jk...}$ in equation~(\ref{eq:Vfunc}) are made, so as to minimise the sum of squared residuals \cite{03YuCaJe.PH3}
\begin{equation} \label{eq:least_sq}
\begin{aligned}
S= & \sum_nw_n\{E_n^{obs}-E_n^{calc}(f_{jk...}+\Delta f_{jk...})\}^2 \\
    & +k\sum_mw_m\{E_m^{\textit{ab}}-E_m^{ref}(f_{jk...}+\Delta f_{jk...})\}^2
\end{aligned}
\end{equation}
The energies $E_n^{calc}$ are found by diagonalising the matrix representation of the Hamiltonian

\begin{equation}
H=T+V+\Delta V
\end{equation}
\noindent
where $\Delta V$ has the same form as Eq.~(\ref{eq:Vfunc}), except $f_{jk...}$ are replaced by the adjustable parameters $\Delta f_{jk...}$.
In the above equation $w_i$ are the weights applied, $E_n^{obs}$ are the experimentally derived energies, and $E_m^{\textit{ab}}$ and $E_m^{\textit{ref}}$ are the energies of the \textit{ab initio} and refined PESs when evaluated on our grid of nuclear geometries.
The second term in Eq.~(\ref{eq:least_sq}) ensures our refined potential retains the general shape of the \textit{ab initio} surface, how strongly we force it to do so is controlled by the constant $k$. In order to find the set of parameters $\Delta f_{jk...}$ for which the above function is minimised, we employ an iterative least-squares fitting algorithm, which requires only the energies and their derivatives with respect to the adjustable parameters.

Because As is heavier than N or P, the rotational energies
of AsH$_3$ are more closely spaced than those of NH$_3$ and PH$_3$, and so more
highly populated at room temperature. Particular attention was therefore paid to
optimising the equilibrium bond lengths and bond angles. This optimisation was
performed prior to the refinement by using the hyperfine resolved rotational energies of
Tarrago \etal ~\cite{96TaDaMa.AsH3}, which we averaged using the spin-statistical weights (\{$A_1$,$A_2$,$E$\}$=$\{16,16,16\}), and a Newton-Gauss
style procedure with a step size of $\pm 0.002$ \AA\ and $\pm 0.002$ rad. Although TROVE is capable of computing quadrupole-hyperfine effects \cite{17YaKuxx},
requiring only a quadrupole moment surface and electric field gradient tensor in addition to the PES and DMS, the resulting splittings are small (roughly a few MHz) and so not considered here.

For the full nonlinear least squares refinement we allowed for corrections to harmonic and certain cubic terms in our PES, and used 322 experimentally
derived energies  with
$J\leq6$ compiled from  Refs.~\cite{96TaDaMa.AsH3,93UkChSh.AsH3,93UkMaWi.AsH3,97YaWaZh.AsH3,
98YaLiWa.AsH3,98WaLiWa.AsH3}.
These sampled the following vibrational  bands: the fundamentals
$\nu_1$, $\nu_2$, $\nu_3$, $\nu_4$; overtones $2\nu_1$, $2\nu_2$, $2\nu_3$,
$2\nu_4$, $3\nu_1$, $3\nu_3$; and  combination bands $\nu_1+\nu_3$,
$\nu_2+\nu_4$, $2\nu_1+\nu_3$, $\nu_1+2\nu_3$. Because we could find no
rotationally excited states belonging to the $\nu_2$ and $\nu_4$ bands in the
literature, only their band centres were included in the refinement.The vibrational band centres
measured by Halonen \etal~\cite{92HaHaBu.AsH3} were not included due to the large estimated uncertainty.
The complete list of experimental energies included in the refinement, along with
their assignments, are included in the supplementary material.

Weights of $w_n=0.1$ were distributed to all experimentally derived rovibrational
states except for purely rotational states, which were given weights of
1000.0, and the $2\nu_4^{l=0}$ band of Yang~\etal\cite{97YaWaZh.AsH3}, for which we struggled to
match experimental energies to our calculated energies owing to conflicting
quantum labels, and so gave a weight of 0.0. These were adjusted on-the-fly
using Watson's robust fitting scheme \cite{03Watson.methods}. A scaling factor of
$k=1\times10^{-4}$ was initially applied to the 39~678 \textit{ab initio} points $E_m^{\textit{ab}}$ included in the fit. As
the refinement progressed this was incrementally decreased to $1\times10^{-6}$
so as to reduce the relative contribution of the \textit{ab initio} data. Care
was taken throughout to ensure the refined PES did not deviate substantially
from the \textit{ab initio} surface, and we note that for all \textit{ab initio}
grid points, the energy difference between the refined and geometry optimised
\textit{ab initio} PES's is
less than 10$\%$ that of the \textit{ab initio} PES above its
zero-point energy (ZPE).

Our final fitted PES is called AsH$_3$-CYT18 below; it is presented as subroutine in the supplementary material.

\section{Dipole moment surface} \label{DMS}
\subsection{Electronic structure calculations}
The electric dipole moment is equal to the first derivative of the electronic
energy with respect to an external electric field. This can be approximated
using a numerical finite-difference procedure whereby the dipole moment
$\overline{\mu}$ is related to the electronic energy in the presence of a weak
uniform electric field $\Delta F$ acting along the space fixed $XYZ$ axis

\begin{equation}
\bar\mu_{X}=\frac{\partial E}{\partial F_X}=\frac{E(+\Delta F_X)-E(-\Delta
F_X)}{2\Delta F_X},
\end{equation}
\begin{equation}
\bar\mu_{Y}=\frac{\partial E}{\partial F_Y},
\end{equation}
\begin{equation}
\bar\mu_{Z}=\frac{\partial E}{\partial F_Z}.
\end{equation}
\noindent
Thus to evaluate the dipole moment at a given nuclear geometry requires seven
electronic structure calculations, the seventh being an initial zero-field
calculation. A field strength of 0.002 a.u. was deemed sufficiently small to
accurately approximate the first derivative without approaching numerical noise \cite{jt744}.
As with our PES, electronic structure calculations were carried out at the
CCSD(T)-F12b level of theory with a cc-pVQZ-PP-F12 basis set for the arsenic
atom, and cc-pVQZ-F12 for the hydrogens. Due to the sevenfold increase in
computational demand of the DMS over the PES, dipole moments were calculated on
a reduced grid of 10,000 points, generated by randomly sampling our PES grid.

\subsection{Analytic representation}
In a similar spirit to our PES, the \textit{ab initio} DMS was  expressed analytically using the symmetrized
molecular bond (SMB) representation\cite{jt466}. In this representation the electronically averaged dipole moment $\bar{\bf{\mu}}$ is constructed as symmetrized projections onto the molecular bonds with the dipole
moment components ($\overline{\mu}_{A_1}$,$\overline{\mu}_{E_a}$,$\overline{\mu}_{E_b}$) in
the molecule fixed axis system given by  4$^{\rm th}$ order polynomial expansions
\begin{equation}
\begin{aligned}
\overline{\mu}_\Gamma(\chi_1,\chi_2,\chi_3,\chi_{4a},\chi_{4b};\rho)= & \mu_0^{\Gamma}(\rho)+\sum_{i}\mu_i^{\Gamma}(\rho)\chi_i + \sum_{i\le j}\mu_{ij}^{\Gamma}(\rho)\chi_i\chi_j\\ + \sum_{i\le j\le k}\mu_{ijk}^{\Gamma}(\rho)\chi_i\chi_j\chi_k
 & + \sum_{i\le j\le k\le l}\mu_{ijkl}^{\Gamma}(\rho)\chi_i\chi_j\chi_k\chi_l ,
\end{aligned}
\end{equation}
where $\Gamma=A_1, E_a$ and $E_b$ are the irreducible components of C$_{\rm 3v}$,
\begin{eqnarray}
\label{eq:mu:icoord1}
\chi_k &=& \Delta r_k \exp(-(\Delta r_k)^2), \quad (k=1,2,3) \\
\label{eq:mu:icoord2}
\chi_{4}&=&(2\alpha_1-\alpha_2-\alpha_3)/\sqrt{6} \\
\chi_{5}&=&(\alpha_2-\alpha_3)/\sqrt{2},
\end{eqnarray}
\begin{equation}
\mu^\Gamma_{ij...}(\rho)=\sum_{s=0}^N\mu^{\Gamma(s)}_{ij...}(\sin(\rho_e)-\sin(\rho))^s,
\end{equation}
$\mu^{\Gamma(s)}_{ij...}$ are the expansion parameters, $\Delta r_k = r_k - r_{\rm eq}$ and $\rho$ is the same as  in Eq.~(\ref{eq:Vfunc}).
The dipole moment components ($\overline{\mu}_{E_a},\overline{\mu}_{E_b}$)  are transformed
as linear combinations of each other by the C$_{\rm 3v}$ group operations and so
transform together as $E$-symmetry.
For this reason the parameters
$(\mu^{E_A(s)}_{ij...},\mu^{E_b(s)}_{ij...})$ must be fit together and
$\mu^{A_1(s)}_{ij...}$ are fitted separately. For an extensive discussion on the SMB representation of
the dipole moment function the reader is directed to [\onlinecite{jt466}].

The final fit required 261 parameters and reproduced the \textit{ab initio} data
with an RMS difference of 0.0008 Debye for energies up to 12,000 cm$^{-1}$,
which is comparable to the level of numerical noise in the finite differences
procedure. The DMS expansion parameter set and a Fortran 90 routine to construct
the DMS is included in the supplementary material.

\section{Results} \label{results}
\subsection{Structural parameters} \label{struc}
Table \ref{table:struc_params} shows the various structural parameters of
AsH$_3$ computed at different levels of theory, compared to those of our
refined PES and those derived from experiment. \textit{Ab initio} calculations
of the equilibrium values of $r$ and $\alpha$ were performed using the geometry
optimisation procedure in MOLPRO. Both VQZ-PP-F12 and AVQZ-DK level calculations
are seen to somewhat overestimate $r_{\rm eq}$ and $\alpha_{\rm eq}$ when compared
experiment, a feature that is exacerbated by the exclusion of relativistic
effects altogether ($r_{\rm eq}^{\rm AVQZ}=1.52375$ \AA, $\alpha_{\rm eq}^{\rm
AVQZ}=92.5553^\circ$ and $r_{\rm eq}^{\rm AV5Z}=1.523653$ \AA, $\alpha_{\rm eq}^{\rm
AV5Z}=92.54910^\circ$). As expected, the effect of our equilibrium geometry
adjustment results in equilibrium bond lengths
and angles much closer to that of experiment \cite{83CaDiFu.AsH3}. This is
reflected in the good agreement between our purely rotational energies and
spin-statistics averaged hyperfine resolved rotational energies of Tarrago \etal~\cite{96TaDaMa.AsH3} (see Table \ref{table:rot}). There are small systematic residuals as large as 0.01 cm$^{-1}$, suggesting our treatment of the rotational motion could be improved by
further tweaking the equilibrium parameters. However, doing so would undoubtedly spoil the vibrational accuracy so we decided against it.

\begin{table*}[h!]
\caption{Experimental and predicted structural constants of $^{75}$AsH$_3$.}
\footnotesize
\begin{tabular}{c c c c c c c c}
\hline
                   & AsH$_3$-CYT18 & VQZ-PP-F12 & AVQZ-DK & AV5Z-DK  &  Exp \cite{83CaDiFu.AsH3}  \\ \hline
$r_{\rm eq}$  /\AA         & 1.511394  &  1.520269 &  1.521481 &  1.520432 & 1.511060 &   & \\
$\alpha_{\rm eq}$  /$^\circ$    & 92.04025  &  92.21595  & 92.17049  & 92.18705 & 92.0690 &   & \\
$r_{\mathrm{SP}}$ /\AA & 1.4688  & 1.4663 & 1.4670 &  &          & \\
$\Delta E$(barrier) / cm$^{-1}$& 14495.  &  14187.  &  14171.  &   &          & \\
\hline	
\end{tabular}
\label{table:struc_params}
\end{table*}

As yet, the inversion barrier height $\Delta E$(barrier) of AsH$_3$
remains unmeasured. The previous highest-level predictions are those
by Schwerdtfeger \etal~\cite{92ScLaPy.AsH3} in 1992, who calculated a
value of 13079.3~cm$^{-1}$ at the Moeller-Plesset (MP2) level of
theory. This is somewhat lower than our CCSD(T) values of just over
14~000 cm$^{-1}$, shown in Table \ref{table:struc_params}. The minimum
energy path over the barrier reduces the As-H bond lengths to their
so-called saddle-point value $r_{\mathrm{SP}}$ at planar geometry. Of
this, the predicted value of 1.457 \AA~ by Schwerdtfeger \etal~is in
reasonable agreement with our own (see Table
\ref{table:struc_params}). For comparison, the NH$_3$ barrier height
is measured to be 1786.8 cm$^{-1}$ occurring for
$r_{\mathrm{SP}}=0.99460$ \AA ~\cite{04RaKaNo.NH3}, and for PH$_3$ the
calculated values of Sousa-Silva \etal~\cite{jt658} are currently the
most reliable, predicting a value of 11~130 cm$^{-1}$ at 1.3611 \AA.
Whereas the NH$_3$ inversion splitting is well known to be
$\approx0.79$ cm$^{-1}$ for the ground vibrational state, it has been
predicted but not observed in PH$_3$ \cite{jt658,18OkSaxx.PH3} and so
it is unlikely to be observed in AsH$_3$ for some time.

\begin{table}[h!]
\centering
\caption{Differences between calculated rotational energies, in cm$^{-1}$, and the hyperfine resolved values of \cite{96TaDaMa.AsH3} which we averaged using the spin statistical weights.}
  \begin{tabular}{ c  c  c  r r}
    \hline
J   &   K   &  Sym  &   Obs         &    Obs$-$Calc$_\mathrm{ref}$\\ \hline
1	&	0	&	A$_2$	&	7.503018	&	-0.000368 \\
1	&	1	&	E	&	7.249824	&	-0.000381 \\
2	&	1	&	E	&	22.253842	&	-0.001287 \\
2	&	2	&	E	&	21.494930	&	-0.000807 \\
3	&	0	&	A$_2$	&	45.005427	&	-0.003161 \\
3	&	1	&	E	&	44.753718	&	-0.002935 \\
3	&	2	&	E	&	43.997254	&	-0.002248 \\
3	&	3	&	A$_1$	&	42.732027	&	-0.001078 \\
3	&	3	&	A$_2$	&	42.732025	&	-0.001077 \\
4	&	1	&	E	&	74.742676	&	-0.005085 \\
4	&	2	&	E	&	73.989211	&	-0.004375 \\
4	&	3	&	A$_1$	&	72.729005	&	-0.003149 \\
4	&	3	&	A$_2$	&	72.728988	&	-0.003184 \\
4	&	4	&	E	&	70.955354	&	-0.001422 \\
5	&	0	&	A$_2$	&	112.460919	&	-0.007935 \\
5	&	1	&	E	&	112.211432	&	-0.007705 \\
5	&	2	&	E	&	111.461647	&	-0.007010 \\
5	&	3	&	A$_1$	&	110.207627	&	-0.005823 \\
5	&	3	&	A$_2$	&	110.207557	&	-0.005824 \\
5	&	4	&	E	&	108.442621	&	-0.004109 \\
5	&	5	&	E	&	106.157387	&	-0.001816 \\
6	&	1	&	E	&	157.148421	&	-0.010773 \\
6	&	2	&	E	&	156.403009	&	-0.010103 \\
6	&	3	&	A$_1$	&	155.156374	&	-0.008751 \\
6	&	3	&	A$_2$	&	155.156164	&	-0.009170 \\
6	&	4	&	E	&	153.401598	&	-0.007307 \\
6	&	5	&	E	&	151.129706	&	-0.005092 \\
    \hline
  \end{tabular}
\label{table:rot}
\end{table}

\subsection{Rovibrational energies} \label{rovib}
Rovibrational energy level calculations were performed up to $J=30$
using the AsH$_3$-CYT18 PES in conjunction with the nuclear motion
program TROVE. Model input parameters were kept the same as reported
in section \ref{nucl_motion}, including our P$_{\mathrm{max}}=14$
vibrational basis. With a basis set of this size the vibrational
Hamiltonian E-symmetry block has 2571 roots. Therefore, given the
2$J$+1 multiplication factor for rotationally exited states, it was
necessary to perform additional basis set truncations to reduce
computational cost. Firstly, our purely vibrational energies $E^i_{\rm
  vib}$ were truncated at 12~000 cm$^{-1}$. These, upon multiplication
with rigid symmetric rotor wavefunctions, form the basis for our full
rovibrational calculation, which we term the $(J=0)$-contracted basis.
Our second truncation, performed only once $J$ exceeded 21, is
therefore to remove all $(J=0)$-contracted eigenfunctions with energy
greater than $E^i_{\rm vib}+E^i_{\rm rotor}=$16~000 cm$^{-1}$, where
$E^i_{\rm rotor}$ are eigenvalues of a symmetric rigid rotor.

Our complete list of calculated energies is available from the ExoMol website (www.exomol.com), along with associated local mode quantum labels
$(n_1,n_2,n_3,n_4,n_5,n_6,\Gamma_{\mathrm{vib}},J,K,\Gamma_{\mathrm{rot}},
\Gamma_{\mathrm{tot}})$. Here $(n_1,n_2,n_3)$ are stretching quantum numbers,
$(n_4,n_5,n_6)$ are bending quantum numbers, $K$ is the projection of the total
rotational angular momentum $J$ onto the molecular axis of symmetry, and
$(\Gamma_{\mathrm{vib}},\Gamma_{\mathrm{rot}},\Gamma_{\mathrm{tot}})$ are the
vibrational, rotational and total symmetry in C$_{\rm 3v}$. The local mode vibrational quantum numbers can be converted to the normal mode representation using symmetry rules
(see Down \etal~\cite{jt538}),
under the assumption that the total number of stretching and bending quanta are conserved between representations. Included in the supplementary material is a list of calculated vibrational states that have been converted to the normal mode representation for $n_1+n_2+n_3\leq 4$ and $n_4+n_5+n_6\leq 4$. This covers all strong bands under 7000 cm$^{-1}$, and should aid any future labelling of AsH$_3$ spectra.

\begin{table*}[h!]
  \caption{Agreement between our calculated energy levels and those derived from experiment. All calculations used our refined PES, AsH$_3$-CYT18. $J=0$ comparisons are before employing the EBSC, and $J=1-6$ comparisons are afterwards. Energy units are cm$^{-1}$.}
\begin{tabular}{l c l l l c l c l l l}
  \\ \hline \hline
  \multirow{2}{*}{Band}
  & \multirow{2}{*}{Symmetry}
      & \multicolumn{4}{c}{$J=0$}
      & \multicolumn{3}{c}{$J=1-6$}\\  \cline{3-6} \cline{8-9}
  & & & Obs. & Calc. & Obs.$-$Calc. & & $\sigma_{\rm rms}^{\rm ebsc}$ & & \\  \hline
    $\nu_2$              & A$_1$ & &  906.752  & 906.109   &    0.643 & & $-$ \\
    $\nu_4$              & E     & &  999.225  & 998.833   &    0.393 & & $-$ \\
    $2\nu_2$             & A$_1$ & & 1806.149  & 1806.161  &   -0.012 & & 0.048 \\
    $\nu_2+\nu_4$        & E     & & 1904.115  & 1904.046  &    0.069 & & 0.131 \\
    $2\nu_4^{l=0}$       & A$_1$ & & 1990.998  & 1990.293  &    0.705 & & 0.262 \\
    $2\nu_4^{l=2}$       & E     & & 2003.483  & 1997.315  &    6.168 & & 0.207 \\
    $\nu_1$              & A$_1$ & & 2115.164  & 2114.938  &    0.227 & & 0.027 \\
    $\nu_3$              & E     & & 2126.432  & 2126.102  &    0.330 & & 0.068\\
    $\nu_1+\nu_2$        & A$_1$ & & 3013$^a$  & 3016.531  &   -3.5   & & $-$ \\
    $\nu_1+\nu_4$        & E     & & 3102$^a$  & 3100.438  &    2.4   & & $-$ \\
    $2\nu_1$             & A$_1$ & & 4166.772  & 4166.694  &    0.078 & & 0.067\\
    $\nu_1+\nu_3$        & E     & & 4167.935  & 4167.877  &    0.058 & & 0.059\\
    $2\nu_3^{l=0}$       & A$_1$ & & 4237.700  & 4237.407  &    0.293 & & 0.046\\
    $2\nu_3^{l=2}$       & E     & & 4247.720  & 4247.842  &   -0.122 & & 0.241\\
    $2\nu_1+\nu_2$       & A$_1$ & & 5057$^a$  & 5040.690  & 16.3  & & $-$ \\
    $\nu_1+\nu_2+\nu_3$  & E     & & 5057$^a$  & 5040.799  & 16.2  & & $-$ \\
    $2\nu_1+\nu_4$       & E     & & 5128$^a$  & 5129.956  &   -2.0   & & $-$ \\
    $2\nu_3^0+\nu_2$     & A$_1$ & & 5128$^a$  & 5131.122  &   -3.0   & & $-$ \\
    $\nu_1+\nu_3+\nu_4$  & E     & & 5158$^a$  & 5155.741  &    2.3   & & $-$ \\
    $\nu_1+\nu_3+\nu_4$  & A$_1$ & & 5158$^a$  & 5156.434  &    1.6   & & $-$ \\
    $3\nu_1$             & A$_1$ & & 6136.340  & 6136.846  &   -0.506 & & $-$\\
    $2\nu_1+\nu_3$       & E     & & 6136.330  & 6136.859  &   -0.529 & & $-$\\
    $\nu_1+2\nu_3^{l=0}$ & A$_1$ & & 6275.830  & 6275.814  &    0.017 & & 0.051\\
    $\nu_1+2\nu_3^{l=2}$ & E     & & 6282.350  & 6282.414  &   -0.064 & & 0.049\\
    $3\nu_3^{l=1}$       & E     & & 6294.710  & 6294.695  &    0.015 & & 0.050\\
    $3\nu_3^{l=3}$       & A$_1$ & & 6365.950  & 6365.759  &    0.191 & & 0.030\\
      \hline \hline
\end{tabular}
\mbox{}\\
$^a$ experimental uncertainties of Halonen \etal~\cite{92HaHaBu.AsH3} are estimated to be 2 cm$^{-1}$ or more. \\
\label{table:comp_eners}
\end{table*}

\begin{figure}[h!]
\centering
\begin{subfigure}{0.5\textwidth}
  \centering
  \includegraphics[width=\linewidth]{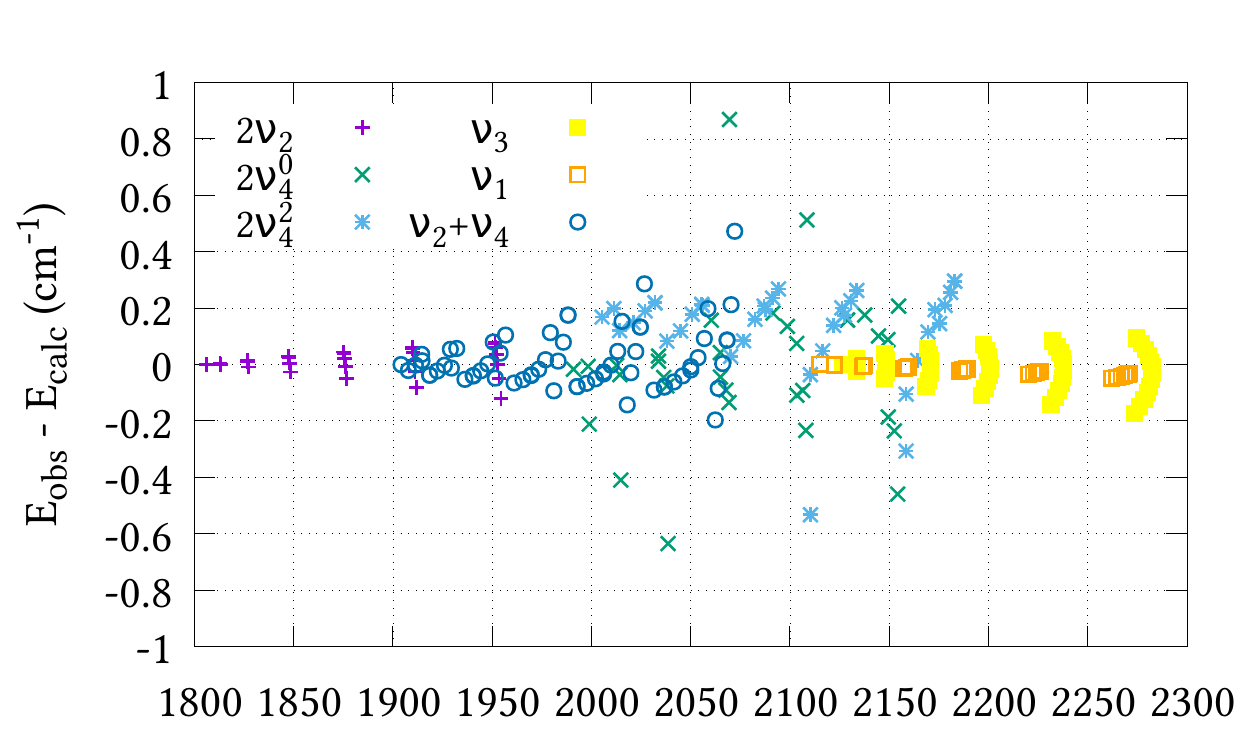}\\
  \includegraphics[width=\linewidth]{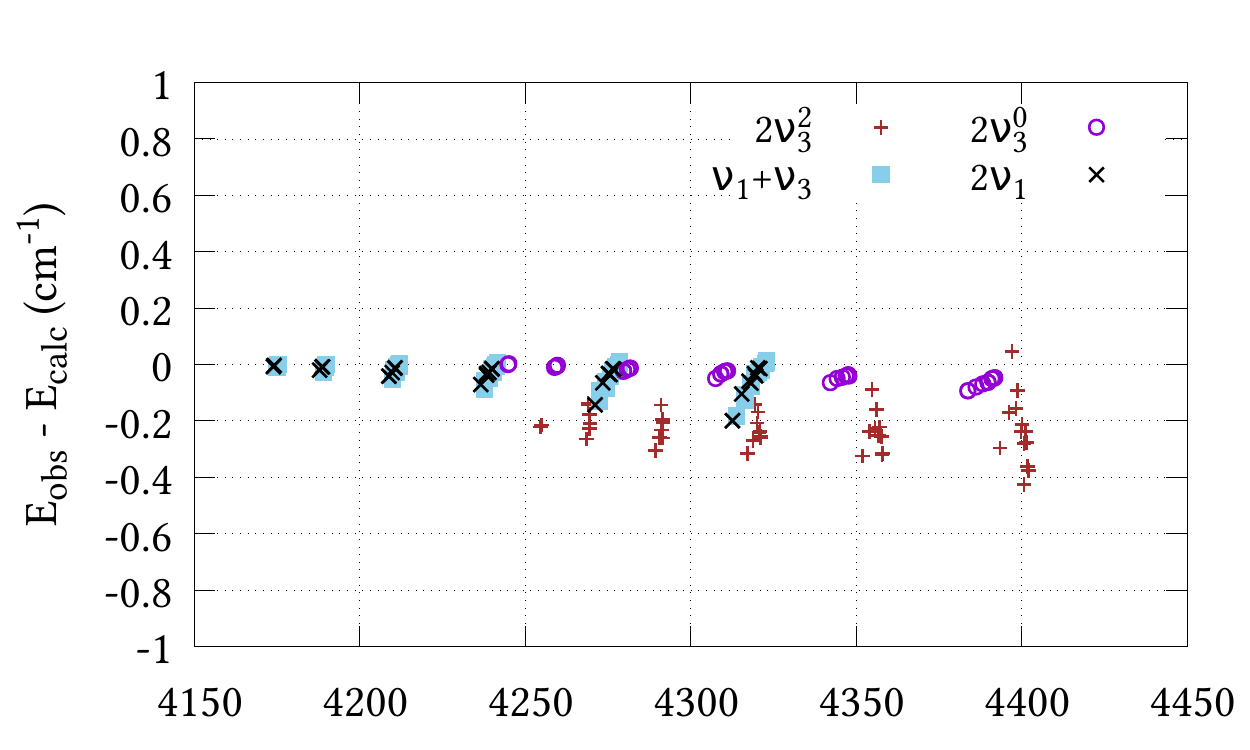}\\
  \includegraphics[width=\linewidth]{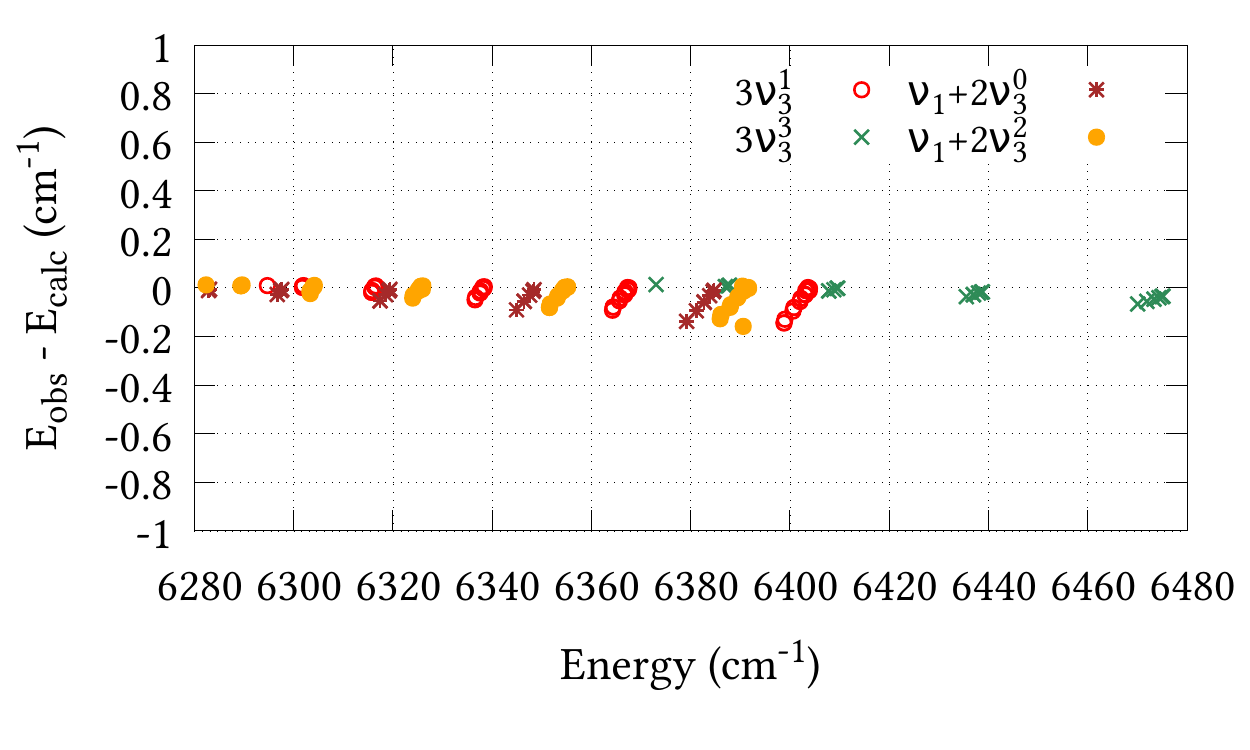}
\end{subfigure}
\caption{Agreement between observed $J=1-6$ energy levels and the calculated values of this work using our refined PES and the EBSC. The $2\nu_2$, $\nu_2+\nu_4$, $2\nu_1$ and $\nu_3$ bands (upper plot) were taken from \cite{93UkChSh.AsH3}; the $2\nu_4^0$ and $2\nu_4^2$ bands (upper plot) were taken from \cite{93UkMaWi.AsH3}; the $2\nu_1$ and $\nu_1+\nu_3$ bands (middle plot) were taken from \cite{97YaWaZh.AsH3}; the $2\nu_3^0$ and $2\nu_3^2$ bands (middle plot) were taken from
\cite{98YaLiWa.AsH3}; and the $3\nu_3^1$, $3\nu_3^3$, $\nu_1+2\nu_3^2$ and $\nu_1+2\nu_3^0$ bands (bottom plot) were taken from reference \cite{98WaLiWa.AsH3}}
\label{fig:AsH3eners}
\end{figure}

Table \ref{table:comp_eners} compares the calculated $J=0$ term values
under 7000 cm$^{-1}$, computed using our refined PES, AsH$_3$-CYT18,
to the experimentally observed values
\cite{08SaLeUl.AsH3,92HaHaBu.AsH3}.  Vibrational labels above 5000
cm$^{-1}$ are tentative.  The refined PES reproduces empirical
energies with a marked improvement over the \textit{ab initio} surface
(see Table \ref{table:p10j0}).  All bands included in the refinement,
except for the $2\nu_4^{l=2}$ band, display sub-wavenumber accuracy.
Based on our energy residuals for $J=1-6$ states belonging to the
$2\nu_4^{l=2}$ band, which fall within $\pm$1.0 cm$^{-1}$ of
experiment, we strongly suspect the empirical band
origin\cite{97YaWaZh.AsH3} of 2003.483 cm$^{-1}$ is incorrect. From a
comparison of our $J=1,2$ energy residuals, we expect the true value
to be closer to 1997.5 cm$^{-1}$. Interestingly, for the bands at 3000
and 5000 cm$^{-1}$ measured by Halonen \etal~\cite{92HaHaBu.AsH3}, all
calculated $J=0$ term values, except for the $2\nu_1+\nu_2$ and
$\nu_1+\nu_2+\nu_3$ bands, fall within a few wavenumbers of experiment
despite being omitted from the refinement. This illustrates the
interpolative power of the refinement, and suggests that even bands
not yet observed experimentally should be predicted with reasonable
accuracy by our refined PES.  Alternative matches for the 5057
cm$^{-1}$ bands within our energies list would be the $\nu_1+3\nu_4^1$
(predicted at 5052.561 cm$^{-1}$) and $\nu_3+3\nu_4^1$ ($A_2$)
(5052.758 cm$^{-1}$) bands. However, considering that these bands are
not predicted to be observable at room temperature, the discrepancies
are more likely due to resonance interactions that are not
well modelled  in our PES.

The residual differences between our calculated $J=0$ term values and those of
experiment can be removed from the final line list by utilising an empirical
basis set correction (EBSC) \cite{jt466}, whereby our calculated band centres are simply
replaced by the corresponding experimental values. We employed the EBSC for all
experimentally known bands taken from \cite{08SaLeUl.AsH3}, except the
suspicious $2\nu_4^{l=2}$ band.
Figure \ref{fig:AsH3eners} displays the difference between our calculated
energies and those derived from experiment for states with $J\leq6$ taken from Refs. \cite{96TaDaMa.AsH3,93UkChSh.AsH3,93UkMaWi.AsH3,97YaWaZh.AsH3,98YaLiWa.AsH3,98WaLiWa.AsH3} after employing the EBSC. The corresponding root-mean-square errors ($\sigma_{\rm rms}^{\rm ebsc}$), split by vibrational band, are shown in table \ref{table:comp_eners}. Although there is some deterioration in quality with $J$, this is slow and systematic in most cases, reassuring us that our calculations can safely be extended to higher rotational excitations. Agreement for the $2\nu_2$ and stretching bands is particularly pleasing, and all calculated $J=1-6$ energies, bar those belonging to the $2\nu_3^2$
band, are calculated to within $\pm0.2$ cm$^{-1}$ of the experimental values. Judging by the systematic offset of the $2\nu_3^2$ band in figure \ref{fig:AsH3eners}, the experimental band center used in the EBSC is most likely $\approx 0.2$ cm$^{-1}$ lower than the true value. Slightly larger $\sigma_{\rm rms}^{\rm ebsc}$ values are observed for the $2\nu_4^0$, $2\nu_4^2$ and $\nu_2+\nu_4$ bands. Whereas the $2\nu_4^2$ and $\nu_2+\nu_4$ bands display clear $J-K$ dependencies, it is difficult to discern any such trends for the $2\nu_4^0$ band, which was omitted from the refinement altogether. Two possible reasons for this are either corrupt experimental data, or perturbation interactions due to nearby states that are not correctly represented by our PES. Even so, the 0.207 cm$^{-1}$ root-mean-square error is very reasonable.

\subsection{Line intensity predictions} \label{eq:intens}
To simulate absolute absorption intensities we use the expression

\begin{equation} \label{eq:intens}
\begin{aligned}
I(f\leftarrow i)= &\frac{8\pi^{3}N_A\nu_{if}\exp(-E''/kT)
[1-\exp(-hc\nu_{if}/kT)]}{(4\pi \epsilon_0)3hcQ} \\
 & \times \sum_{\Phi_{int}'\Phi_{int}''}\sum_{A=X,Y,Z}|
 \langle \Phi_{int}'|\mu_A|\Phi_{int}''\rangle|^2
\end{aligned}
\end{equation}

\noindent
where $\Phi_{int}'$ and $\Phi_{int}''$ are the upper and lower state
wavefunctions respectively, that correspond to energies $E'$ and $E''$.
$\nu_{if}$ is the transition frequency, $\mu_A$ is the electric dipole moment
along the $A=X,Y,Z$ axis, T is the absolute temperature and $Q(T)$ the partition
function given by
\begin{equation} \label{eq:pfunc}
Q=\sum_w g_w \exp(-E_w/kT)
\end{equation}
In the above equation, $E_w$ is the energy and $g_w$ is the total
degeneracy of state $w$. Note that we are using the same symbol $E$
for the energies (Joule) and term values or wavenumbers (cm$^{-1}$),
whereas conventionally the latter should be $\tilde{E}$.

\begin{figure}
\centering
  \includegraphics[width=\linewidth]{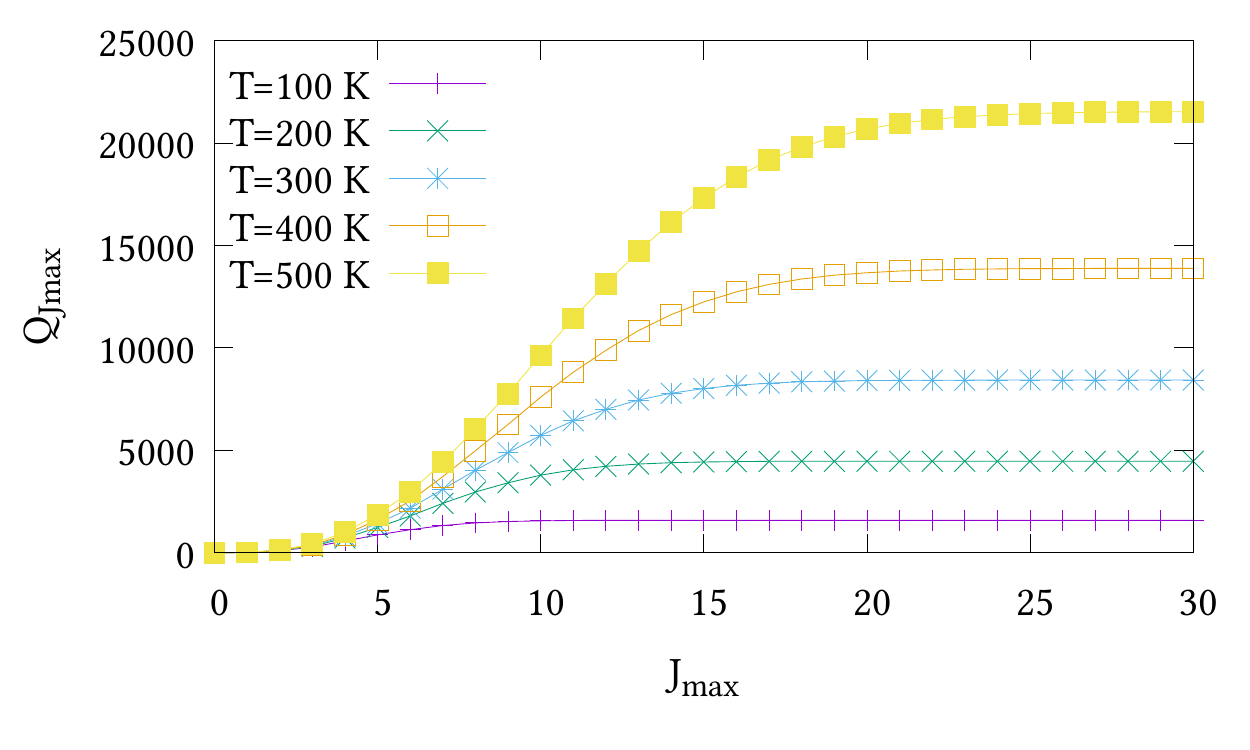}
  \caption{The partition functions $Q_{J_{max}}$ of AsH$_3$ at different temperatures versus the maximum $J$ value used in Eq. (\ref{eq:pfunc})}
  \label{fig:pfuncs}
\end{figure}

The nuclear spin statistical weights for AsH$_3$ are \{16,16,16\} for
states of \{$A_1$,$A_2$,$E$\} symmetry, and so $g_w=16(2J_w+1)$. No
calculated or experimentally derived values of the partition function
could be found in the literature, so we provide partition function
values in the supplementary material, for temperatures ranging from 10
to 500 K in intervals of 10 K. Fig.~\ref{fig:pfuncs} illustrates the
convergence of $Q$ as the rotational basis is increased from including
only $J=0$ states ($J_{\rm max}=0$), to all computed states with
$J\leq30$ ($J_{\rm max}=30$). In reality there will be additional
contributions from our vibrational basis (P$_{\rm max}=14$) and PES,
although these are difficult to quantify. The room temperature
partition function was calculated to be $Q(T=296)=8250.2801$ using
$J_{\rm max}=30$, which we estimate to be better than 99\% converged.

\begin{figure}
\centering
  \includegraphics[width=\linewidth]{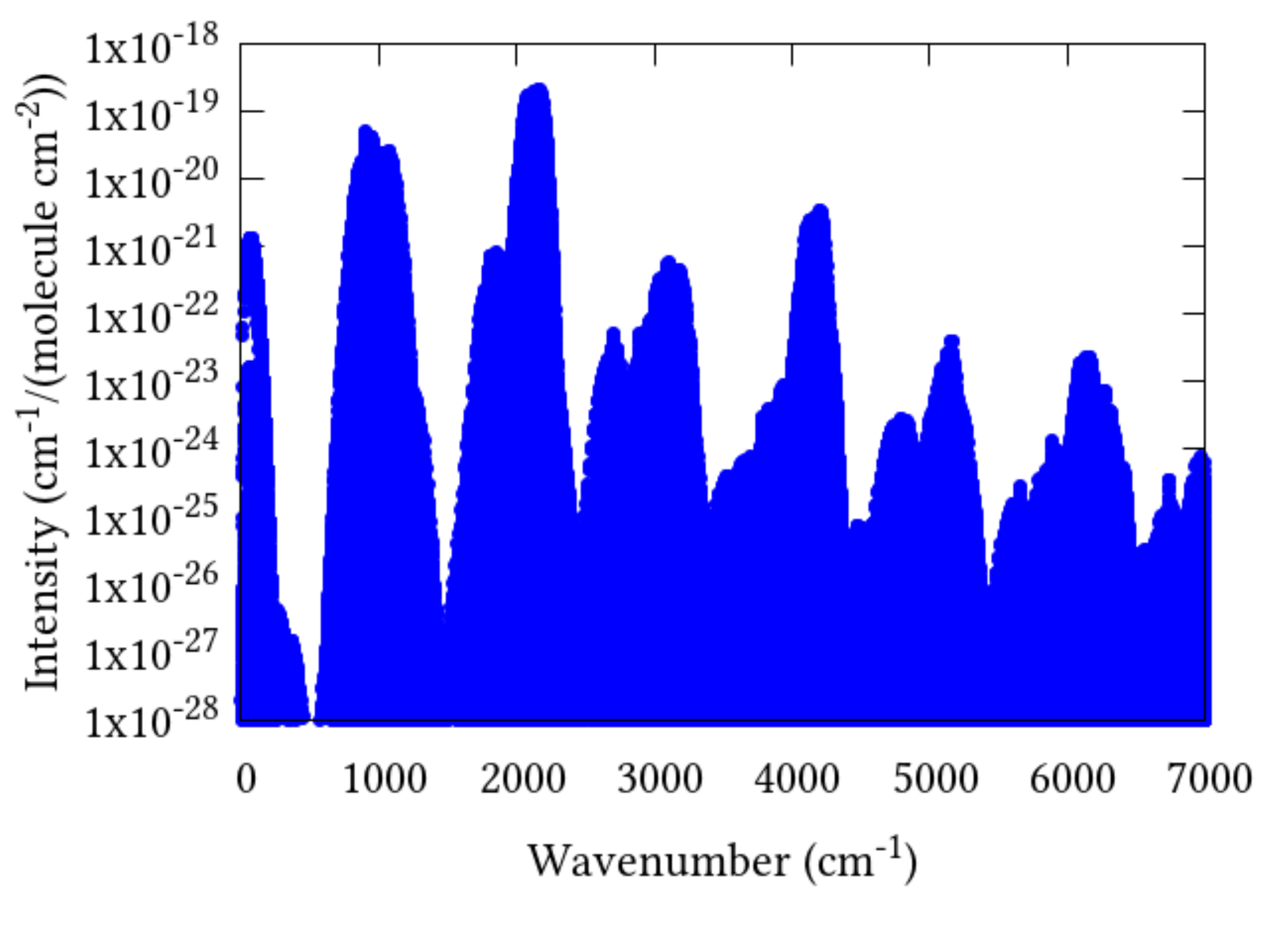}
  \caption{Overview of complete $J=0-30$ line list computed at 296 K.}
  \label{fig:overview}
\end{figure}

\begin{table*}[!h]
\centering
\caption{Comparison of observed and calculated band intensities. Column 1 refers to the local mode quantum numbers assigned by TROVE, where sym is the total symmetry. The units of intensity are 10$^{-18}$ cm$^{-1}/(\mathrm{molec}$ cm$^{-2}$). The value under / is the total intensity of the bands with the same quantum numbers $n_1n_2n_3$}
 \begin{tabular}{l  l  c  l  l  l}
    \hline
($n_1,n_2,n_3$;sym) & band   &  bend centre   &  $I_\mathrm{obs}$\cite{ 02ZhHeDi.AsH3}  &   $I_\mathrm{calc}$\cite{ 02ZhHeDi.AsH3}      & $I_{\mathrm{calc}}$ (this work) \\ \hline
(100;$A_1$)	&$\nu_1$	& 2115.164	& 11.4/44.1 & 10.7/40.4 & 11.2/44.9\\
(100;$E$)	&$\nu_3$	& 2126.432	& 29.7 & 32.7 & 33.7\\
(200;$A_1$)	&$2\nu_1$	& 4166.772	& /0.618 & 0.157/0.427 & 0.225/0.722\\
(200;$E$)	&$\nu_1+\nu_3$	& 4167.935	& $-$ & 0.270 & 0.497 \\
(110;$A_1$)	& $2\nu_3^{l=0}$ & 4237.700	& $-$ & 0.0117/0.0123 & 0.0143/0.0163 \\
(110;$E$)	& $2\nu_3^{l=2}$ & 4247.720	& $-$ & 0.000671 & 0.00201 \\
(300;$A_1$)	& $3\nu_1$ & 6136.340	& /0.00989 & 0.00456/0.00656 & 0.00337/0.00548 \\
(300;$E$)	& $2\nu_1+\nu_3$ & 6136.330	& $-$ & 0.00200 & 0.00211 \\
(210;$A_1$)	& $\nu_1+2\nu_3^{l=0}$ & 6275.830	& /0.00275 & 0.00112/0.00182 & 0.000734/0.00116 \\
(210;$E$)	& $\nu_1+2\nu_3^{l=2}$ & 6282.350	& $-$      & 0.000104        & 0.0000741 \\
(210;$E$)	& $3\nu_3^{l=1}$       & 6294.710	& $-$      & 0.000596        & 0.000356 \\
(111;$A_1$)	& $3\nu_3^{l=3}$       & 6365.950	& $-$      & 0.0000650        & 0.0000934 \\
    \hline
  \end{tabular}
\label{table:intens}
\end{table*}

\begin{table*}
\caption{Comparison of calculated and observed \cite{93DaMaTa.AsH3} line positions and intensities belonging to the $\nu_1$ and $\nu_3$ bands.}
  \begin{tabular}{ c  c  c  l  c  c  c  c  c  c  c  c}
    \hline
$J$' &   $K'$ &  Sym$'$ & $J''$ & $K''$ &  Sym$''$ & band & $\nu_\mathrm{obs}$\cite{93DaMaTa.AsH3} & $I_{\mathrm{obs}}$\cite{93DaMaTa.AsH3} & $\nu_\mathrm{calc}$ & $I_{\mathrm{calc}}$ & \%$|\frac{I_{\mathrm{obs}}-I_{\mathrm{calc}}}{I_{\mathrm{obs}}}|$ \\ \hline		
9	&	6	&	E	&	10	&	7	&	E	&	$\nu_3$	&	2051.894	&	4.799$\times$10$^{-20}$	&	2052.082	&	4.992$\times$10$^{-20}$	&	4.03	\\
9	&	7	&	E	&	10	&	8	&	E	&	$\nu_3$	&	2052.548	&	5.755$\times$10$^{-20}$	&	2052.767	&	6.079$\times$10$^{-20}$	&	5.63	\\
7	&	1	&	A$_2$	&	8	&	0	&	A$_1$	&	$\nu_3$	&	2064.460	&	4.396$\times$10$^{-20}$	&	2064.468	&	4.773$\times$10$^{-20}$	&	8.57	\\
7	&	6	&	E	&	8	&	7	&	E	&	$\nu_3$	&	2067.961	&	1.038$\times$10$^{-19}$	&	2068.139	&	1.110$\times$10$^{-19}$	&	6.95	\\
4	&	4	&	E	&	5	&	5	&	E	&	$\nu_3$	&	2090.433	&	1.572$\times$10$^{-19}$	&	2090.542	&	1.699$\times$10$^{-19}$	&	8.04	\\
4	&	1	&	A$_1$	&	5	&	0	&	A$_2$	&	$\nu_3$	&	2088.098	&	5.876$\times$10$^{-20}$	&	2088.087	&	6.423$\times$10$^{-20}$	&	9.31	\\
5	&	4	&	E	&	6	&	5	&	E	&	$\nu_3$	&	2082.601	&	1.250$\times$10$^{-19}$	&	2082.714	&	1.320$\times$10$^{-19}$	&	5.64	\\
3	&	3	&	E	&	4	&	4	&	E	&	$\nu_3$	&	2097.659	&	1.462$\times$10$^{-19}$	&	2097.738	&	1.571$\times$10$^{-19}$	&	7.49	\\
2	&	2	&	E	&	1	&	1	&	E	&	$\nu_3$	&	2140.716	&	8.659$\times$10$^{-20}$	&	2140.678	&	9.453$\times$10$^{-20}$	&	9.17	\\
2	&	1	&	A$_1$	&	1	&	0	&	A$_2$	&	$\nu_3$	&	2141.069	&	8.949$\times$10$^{-20}$	&	2141.053	&	9.612$\times$10$^{-20}$	&	7.41	\\
6	&	6	&	E	&	5	&	5	&	E	&	$\nu_3$	&	2168.331	&	1.965$\times$10$^{-19}$	&	2168.331	&	2.093$\times$10$^{-19}$	&	6.48	\\
8	&	5	&	E	&	7	&	5	&	E	&	$\nu_1$	&	2172.196	&	3.114$\times$10$^{-20}$	&	2172.262	&	3.301$\times$10$^{-20}$	&	5.99	\\
8	&	7	&	E	&	9	&	7	&	E	&	$\nu_1$	&	2045.190	&	2.459$\times$10$^{-20}$	&	2045.261	&	2.326$\times$10$^{-20}$	&	5.40	\\
8	&	8	&	E	&	9	&	8	&	E	&	$\nu_1$	&	2045.319	&	1.310$\times$10$^{-20}$	&	2045.397	&	1.328$\times$10$^{-20}$	&	1.36	\\
10	&	7	&	E	&	9	&	7	&	E	&	$\nu_1$	&	2185.605	&	2.026$\times$10$^{-20}$	&	2185.716	&	2.114$\times$10$^{-20}$	&	4.35	\\
    \hline
  \end{tabular}
\label{table:intens2}
\end{table*}

Line list calculations were performed using the AsH$_3$-CYT18 PES and the cc-pVQZ-PP-F12 DMS detailed in section \ref{DMS}. Transitions involve states with energies up to 10~500 cm$^{-1}$, rotational excitation up to $J=30$, and a maximum lower state energy of 3500 cm$^{-1}$. The final line list consists of 3.6 million absorption lines in the range $0-7000$ cm$^{-1}$ with intensity greater than 1$\times10^{-28}$ cm$^{-1}$/(molecule cm$^{-2}$) at 296 K. An overview is presented in figure \ref{fig:overview}.
It is available to download from the ExoMol website (www.exomol.com), where it is provided in the ExoMol  format \cite{jt631}.
Summarily, this consists of a .states file which contains the complete list of rovibrational states with associated energies and quantum numbers,  and a .trans file which contains the complete list of transitions, each identified by an upper and lower state index (in the .states file), Einstein A-coefficient, and transition wavenumber.

Several sources of experimental absorption data exist for AsH$_3$. In the following paragraphs our intensity calculations are validated by comparison with only the most recent and reliable sources. For the first test of our absolute intensities we compare our calculated band
intensities with those obtained by Zheng \etal~\cite{ 02ZhHeDi.AsH3}, shown in Table
\ref{table:intens}. Zheng \etal\ produced a three-dimensional DMS based on of density functional theory calculations, and compared the resulting absolute vibrational band intensities to the values obtained by direct integration of absorbance spectra, which they provide with 20$-$40\% estimated uncertainty. Due to multiple bands overlapping only the combined intensity of bands with the same local mode quanta are presented in some cases. For the $\nu_1$ and $\nu_3$ fundamentals we compare well
with experiment, reproducing the observed values within 2\% and 14\%
respectively. Zheng \etal\ only provide the measured intensity of the sum of the
$2\nu_1$ and $\nu_1+\nu_3$ bands, for which we are stronger by 17\%. No measurements of the weaker  $2\nu_3^{l=0}$ and $2\nu_3^{l=2}$ bands are given, most likely due to difficulties resolving them without accurate theoretical line positions. Finally, for the three-quanta stretches, our calculated intensities are typically 2-3 times weaker than the measured values. However, it is difficult to estimate the reliability of these measurements, given the recorded spectrum is only medium-resolution ($\Delta \nu=0.2$ cm$^{-1}$) and the bands are weak.

Dana \etal~\cite{93DaMaTa.AsH3} measured absolute intensities of 387 lines belonging
to the $\nu_1$ and $\nu_3$ bands. Their line measurements range from 2010$-$2235
cm$^{-1}$ although they make no attempt to measure the $Q-$branch from
2110$-$2140 cm$^{-1}$, presumably owing to the density of lines. Table
\ref{table:intens2} compares our calculated line positions and intensities to
the experimentally measured values for 14 randomly selected strong lines
measured by Dana \etal\ In all cases our calculated intensity values are within
$\pm$10\% of experiment, although there is a slight tendency to be higher. Nevertheless, this is reassuring given our 14\% discrepancy with
the $\nu_3$ band as measured by Zheng \etal~\cite{ 02ZhHeDi.AsH3}.

\begin{figure*}[!h]
\centering
  \includegraphics[width=0.7\linewidth]{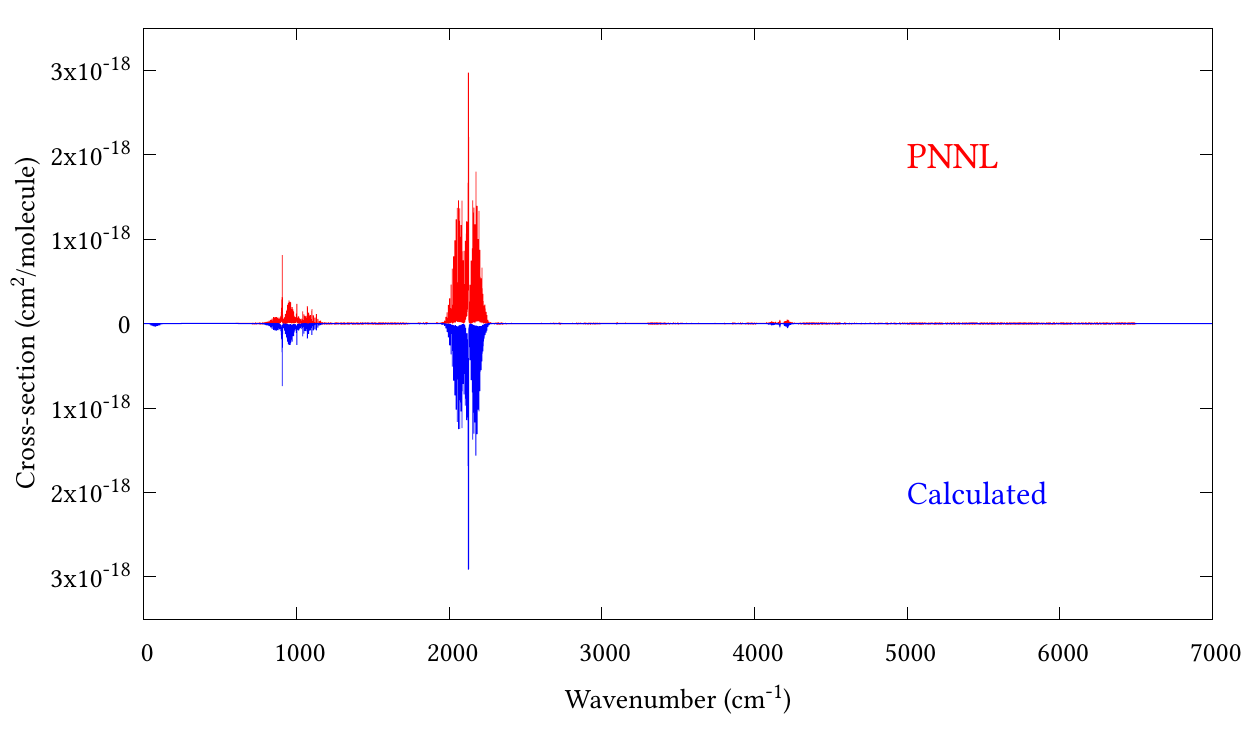}
  \caption{Overview of synthetic $J=0-30$ spectrum computed at 298.15 K compared to PNNL. 0$-$7000 cm$^{-1}$ region.}
  \label{fig:3}
\end{figure*}

The PNNL database provides a composite spectrum of pure AsH$_3$ up to 6500 cm$^{-1}$
measured at 5, 25 and 50$^{\circ}$C. For comparison, we generated synthetic
T=298.15 K spectra using a $J=0-30$ line list convoluted with a Voigt
profile with half-width at half-maxima (HWHM) of 0.09 cm$^{-1}$. Although linewidths are well known to depend upon the upper and lower state quantum numbers, the strongest dependency being $J$ and $K$, as far as we know no such data exists for AsH$_3$, and we found the value 0.09 cm$^{-1}$ reasonably approximated the PNNL linewidths on average. To convert the PNNL
absorbance spectra to cm$^2$/molecule a multiplication factor of 9.28697$\times 10^{-16}$ is necessary.

\begin{figure}[!h]
\centering
  \includegraphics[width=\linewidth]{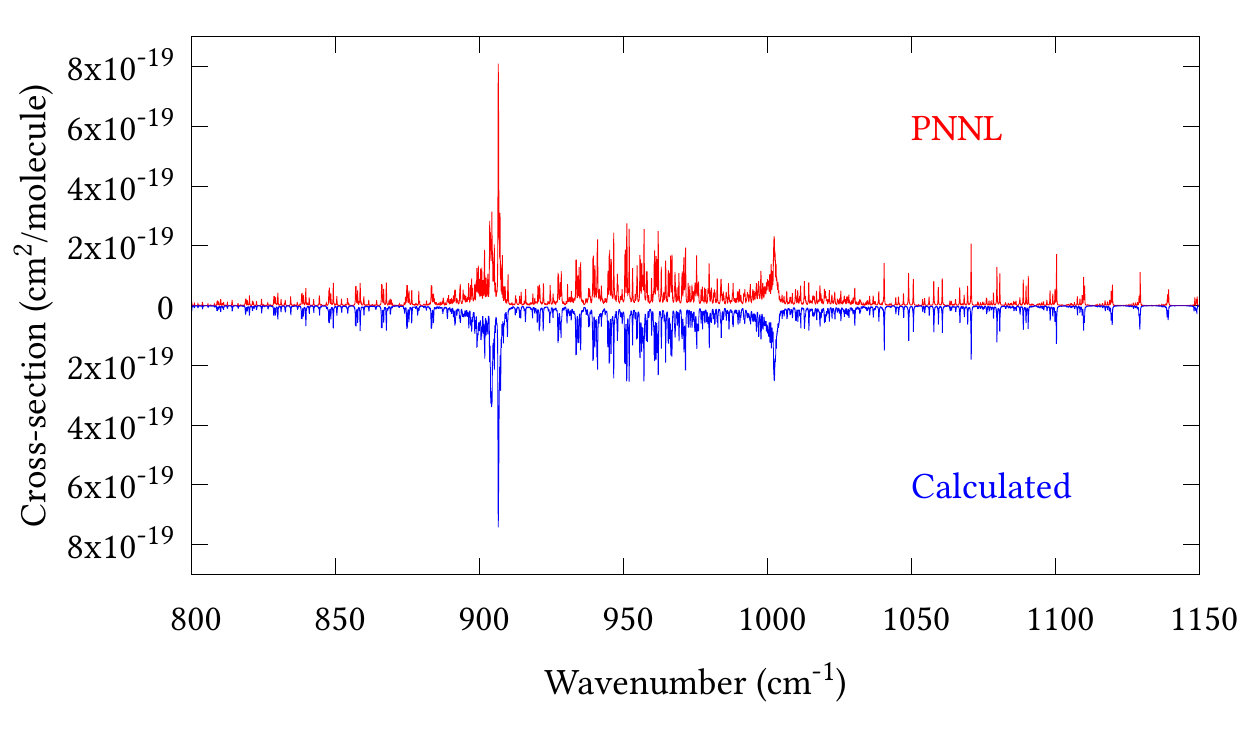}
  \caption{Expansion of synthetic $J=0-30$ spectrum computed at 298.15 K compared to PNNL. 800$-$1150 cm$^{-1}$ region.}
  \label{fig:4}
\end{figure}

\begin{figure}[!h]
\centering
  \includegraphics[width=\linewidth]{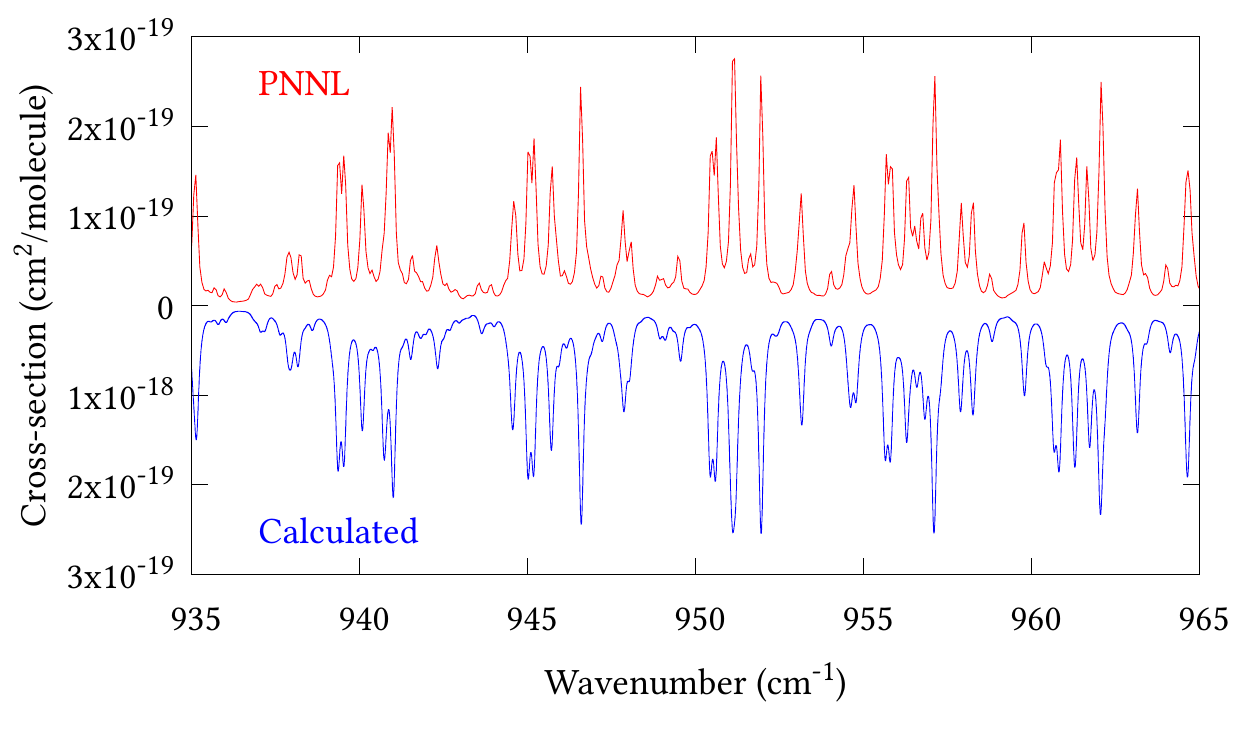}
  \caption{Expansion of synthetic $J=0-30$ spectrum computed at 298.15 K compared to PNNL. 935$-$965 cm$^{-1}$ region.}
  \label{fig:4a}
\end{figure}

\begin{figure}[!h]
\centering
  \includegraphics[width=\linewidth]{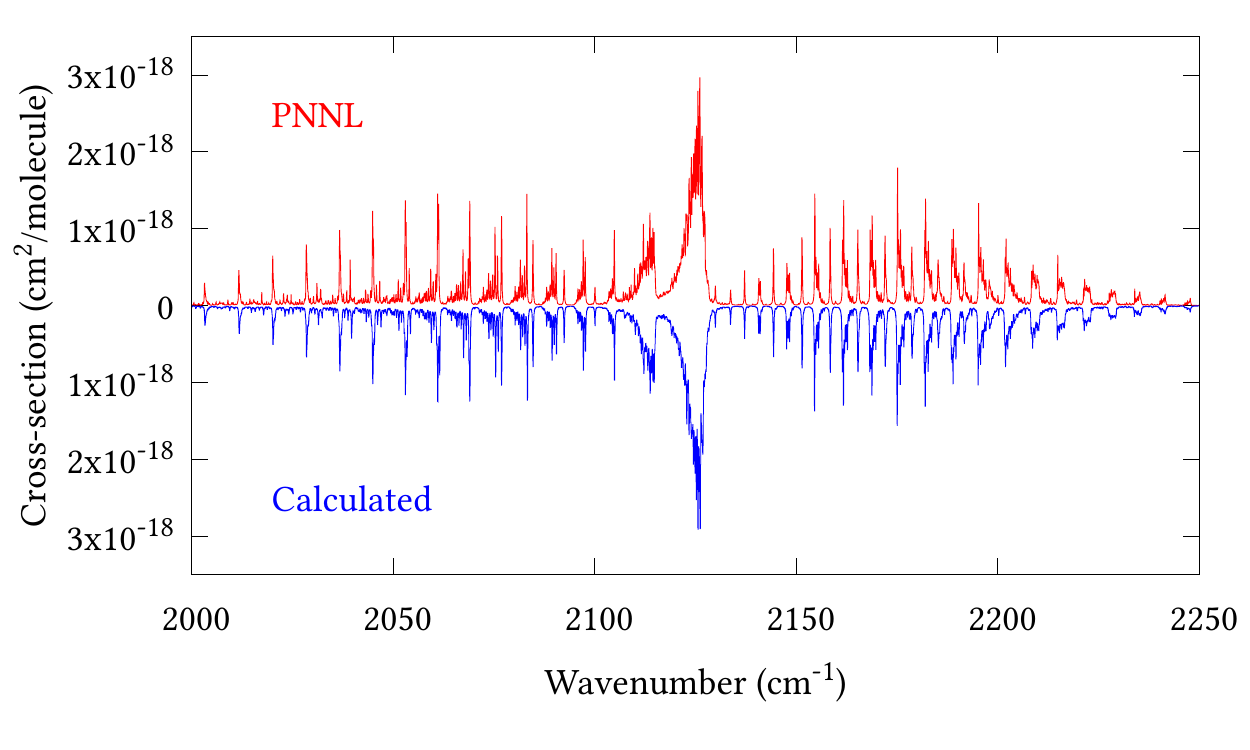}
  \caption{Expansion of synthetic $J=0-30$ spectrum computed at 298.15 K compared to PNNL. 2000$-$2250 cm$^{-1}$ region.}
  \label{fig:5}
\end{figure}

\begin{figure}[!h]
\centering
  \includegraphics[width=\linewidth]{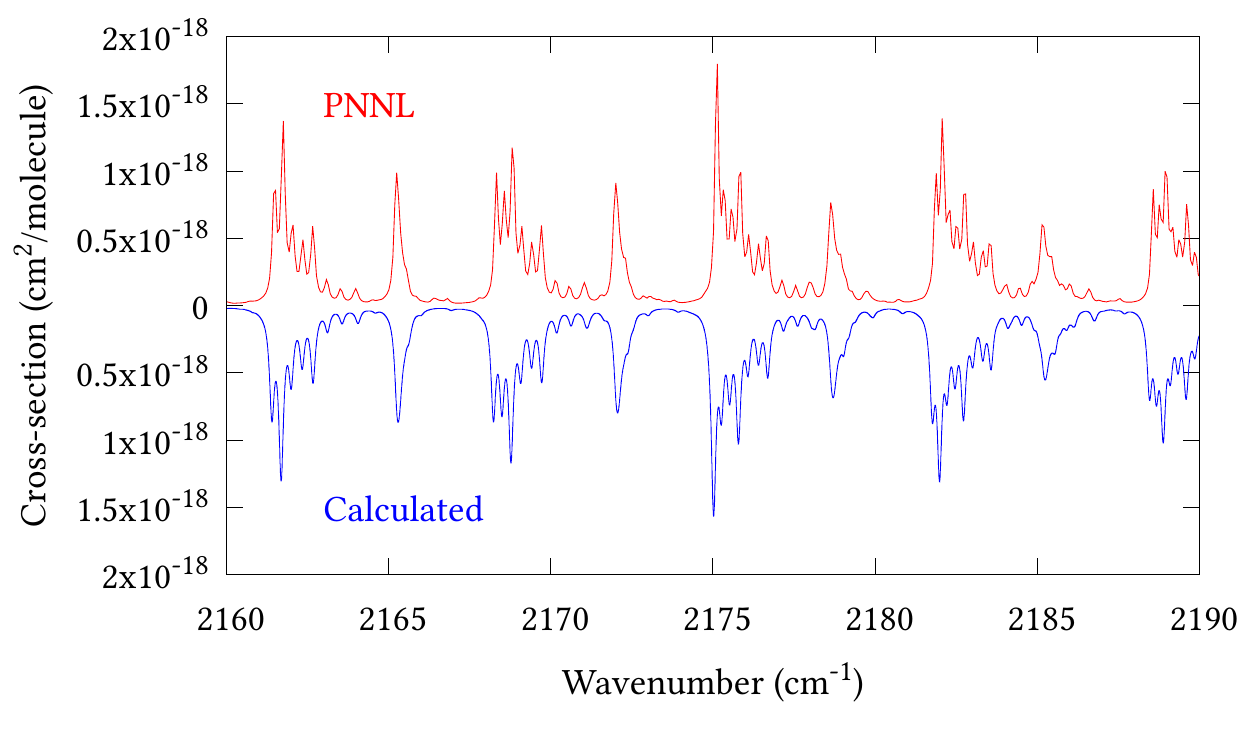}
  \caption{Expansion of synthetic $J=0-30$ spectrum computed at 298.15 K compared to PNNL. 2160$-$2190 cm$^{-1}$ region.}
  \label{fig:5a}
\end{figure}

\begin{figure}[!h]
\centering
  \includegraphics[width=\linewidth]{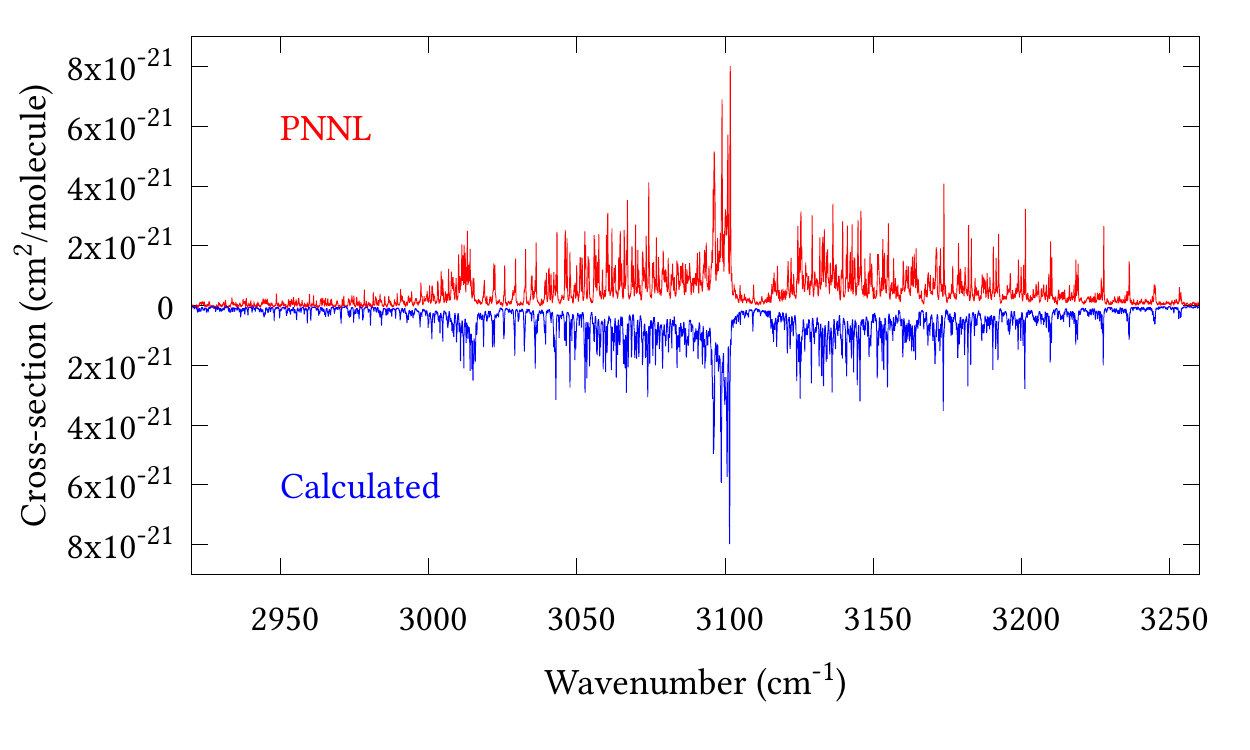}
  \caption{Expansion of synthetic $J=0-30$ spectrum computed at 298.15 K compared to PNNL. 2920$-$3260 cm$^{-1}$ region.}
  \label{fig:6}
\end{figure}

\begin{figure}[!h]
\centering
  \includegraphics[width=\linewidth]{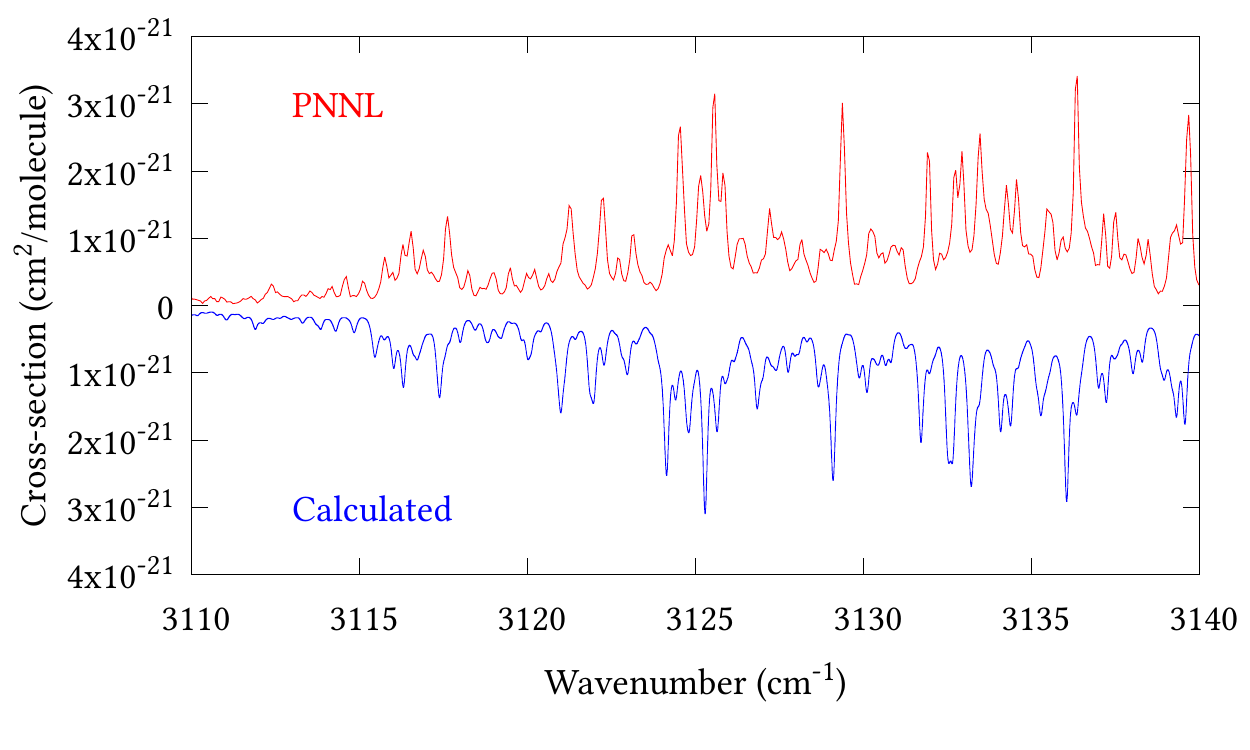}
  \caption{Expansion of synthetic $J=0-30$ spectrum computed at 298.15 K compared to PNNL. 3110$-$3140 cm$^{-1}$ region.}
  \label{fig:6a}
\end{figure}

\begin{figure}[!h]
\centering
  \includegraphics[width=\linewidth]{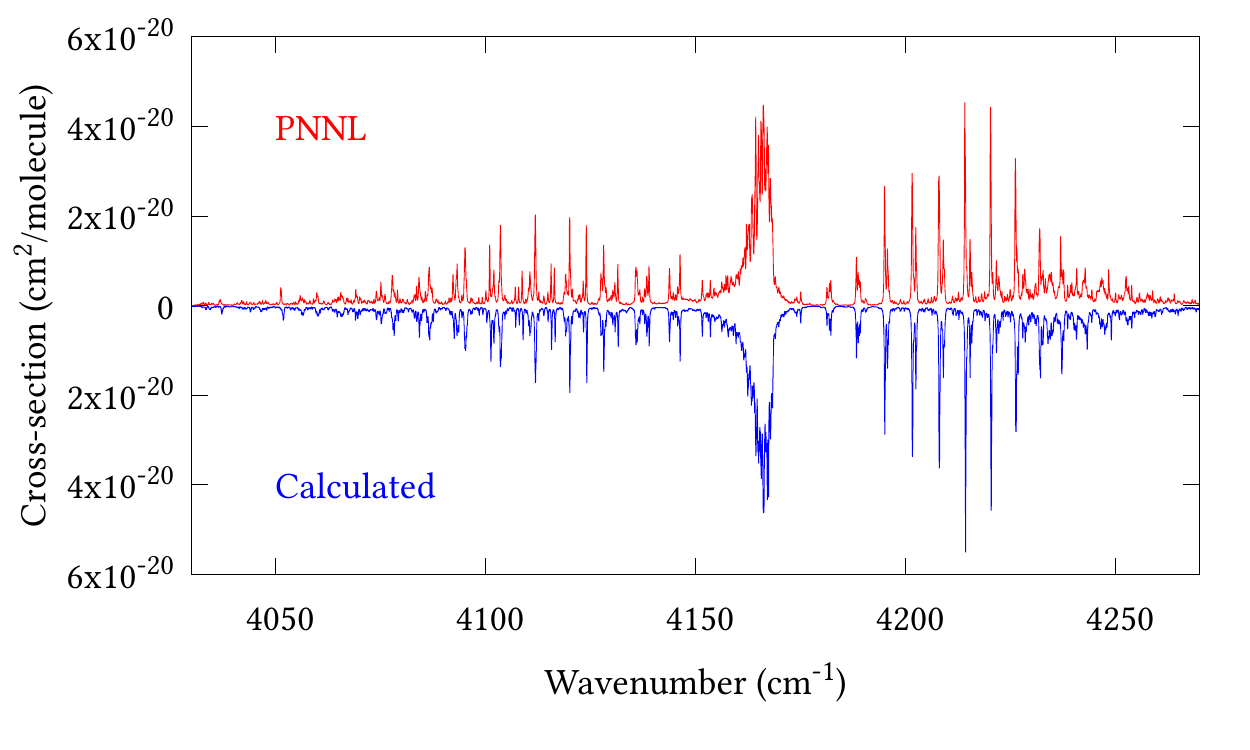}
  \caption{Expansion of synthetic $J=0-30$ spectrum computed at 298.15 K compared to PNNL. 4035$-$4285 cm$^{-1}$ region.}
  \label{fig:7}
\end{figure}

\begin{figure}[!h]
\centering
  \includegraphics[width=\linewidth]{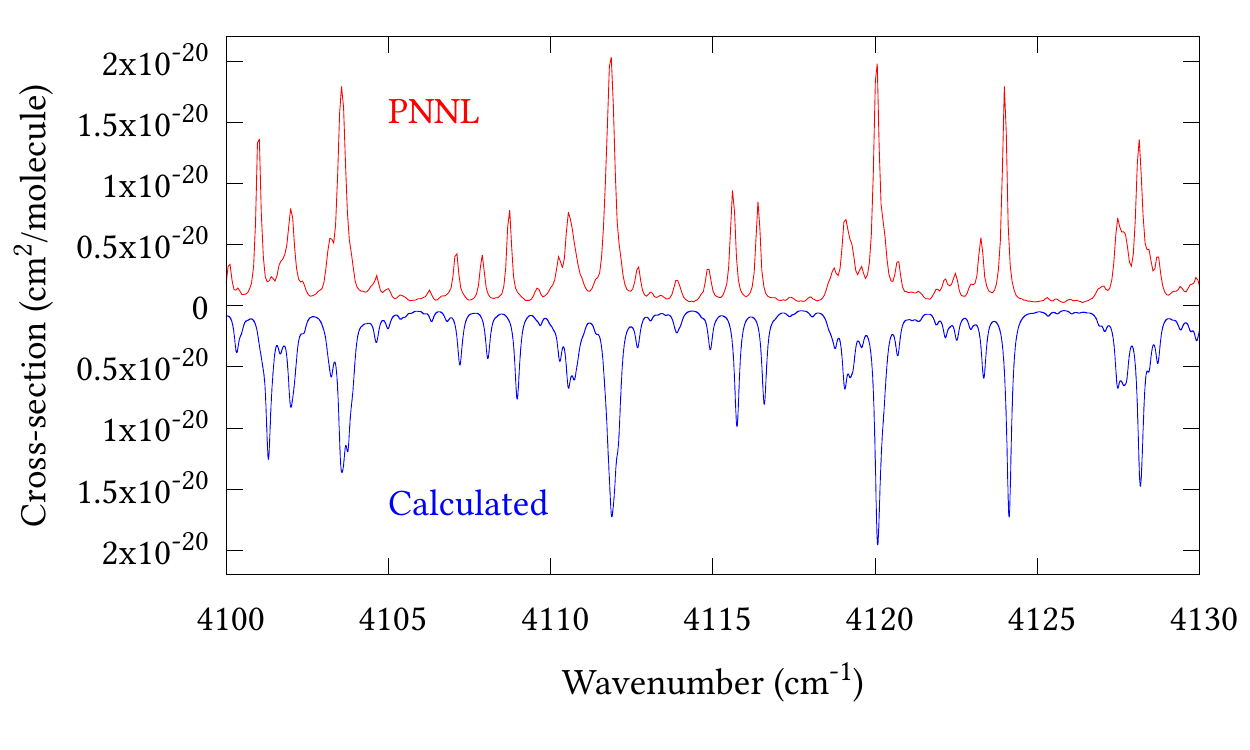}
  \caption{Expansion of synthetic $J=0-30$ spectrum computed at 298.15 K compared to PNNL. 4100$-$4130 cm$^{-1}$ region.}
  \label{fig:7a}
\end{figure}

\begin{figure}[!h]
\centering
  \includegraphics[width=\linewidth]{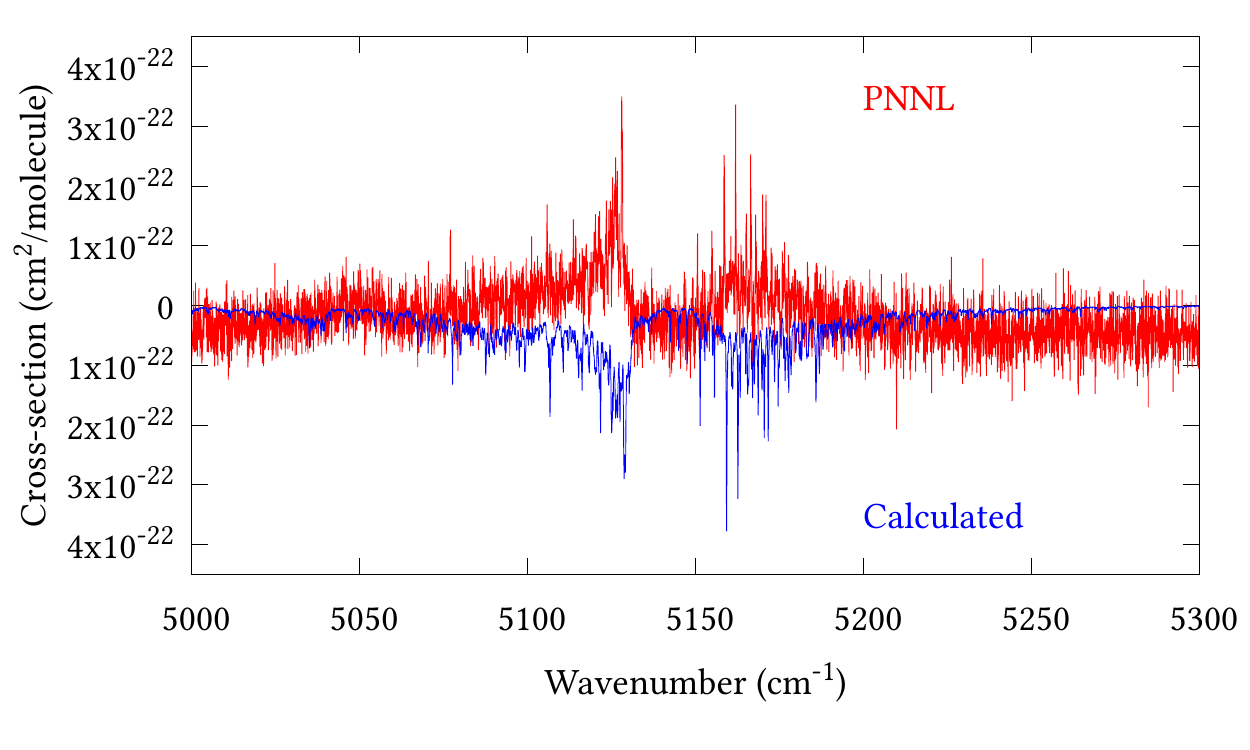}
  \caption{Expansion of synthetic $J=0-30$ spectrum computed at 298.15 K compared to PNNL. 5000$-$5300 cm$^{-1}$ region.}
  \label{fig:8}
\end{figure}

\begin{figure}[!h]
\centering
  \includegraphics[width=\linewidth]{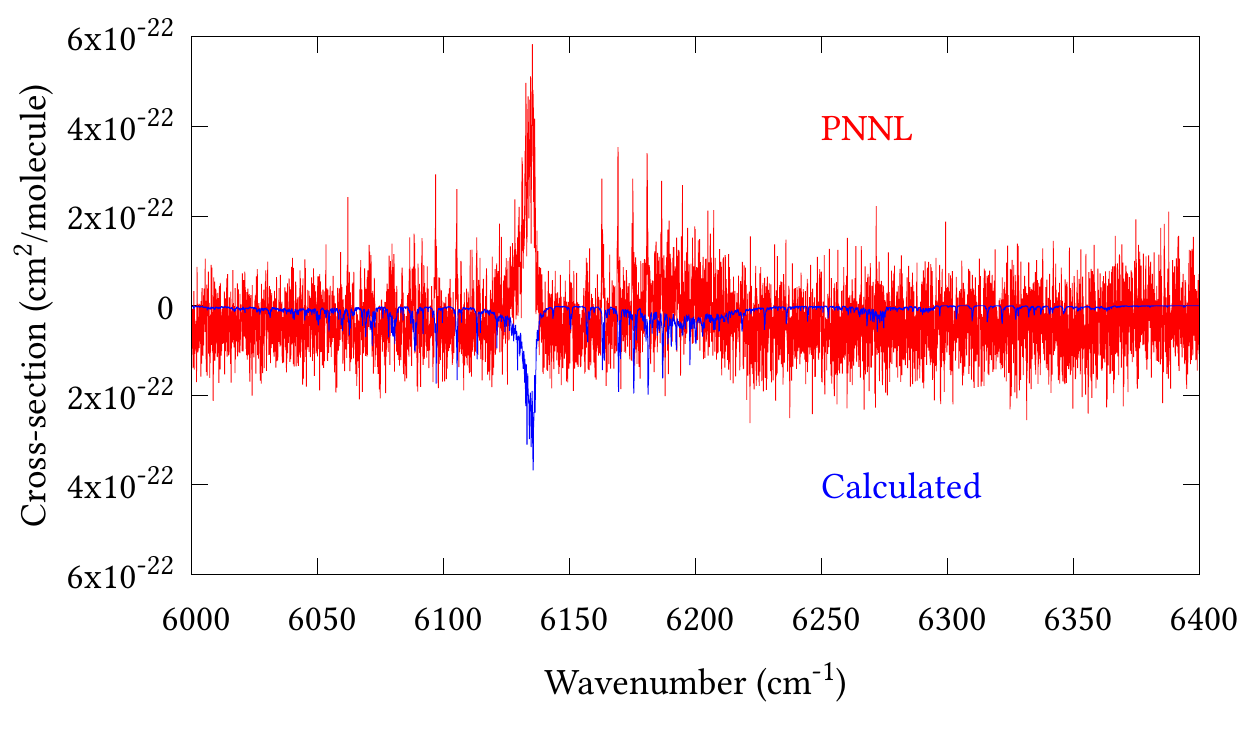}
  \caption{Expansion of synthetic $J=0-30$ spectrum computed at 298.15 K compared to PNNL. 6000$-$6400 cm$^{-1}$ region.}
  \label{fig:9}
\end{figure}

Figure \ref{fig:3} shows an overview of our synthetic spectrum compared to PNNL,
with the key absorption features expanded in Figs. \ref{fig:4}-\ref{fig:9}.
Qualitative agreement is very good, particularly for the $\nu_1$/$\nu_3$ (see
Figs.~\ref{fig:5} and \ref{fig:5a}) and $\nu_2$/$\nu_4$ (see Figs.~\ref{fig:4} and \ref{fig:4a}) fundamentals, and
the $2\nu_1$/$\nu_1+\nu_3$/$2\nu_3^0$/$2\nu_3^2$ band system (shown
Figs.~\ref{fig:7} and \ref{fig:7a}). Note that despite only
including the $\nu_2$ and $\nu_4$ band centres in the refinement, their
rotational structures are reproduced well. In the 2920$-$3260 cm$^{-1}$ region
(shown Figs.\ref{fig:6} and \ref{fig:6a}) the dominant sources of opacity are predicted to be the
strong $\nu_1+\nu_4$ and $\nu_3+\nu_4$ (A$_1$) (calculated band centre 3119.400
cm$^{-1}$) bands, and the slightly weaker $\nu_2+\nu_3$ (3023.706 cm$^{-1}$) and
$\nu_1+\nu_2$ bands. Considering that no associated experimental energies were
included in the refinement, the level of agreement is satisfying. Above 5000
cm$^{-1}$ most absorption features are lost in the PNNL background noise; only
the $2\nu_1+\nu_4$, $\nu_2+2\nu_3^0$ and $\nu_1+\nu_3+\nu_4$ bands (our
labelling) are clearly visible between 5000--5250 cm$^{-1}$ (see Fig. \ref{fig:8}). There is a tenuous absorption bump in PNNL at 5050 cm$^{-1}$ for which we
appear to be offset by roughly 15 cm$^{-1}$, confirming our discrepancies with
the $2\nu_1+\nu_2$ and $\nu_1+\nu_2+\nu_3$
band centres measured by Halonen \etal~\cite{92HaHaBu.AsH3} (see table
\ref{table:comp_eners}). In the $6000-6400$ cm$^{-1}$ region (see Fig.
\ref{fig:9}) the salient feature
is the $3\nu_1$ and $2\nu_1+\nu_3$ Q-branch at 6135 cm$^{-1}$, for which we
clearly underestimate the intensity. Although in line with our comparisons with
Zheng \etal~\cite{ 02ZhHeDi.AsH3} (see table \ref{table:intens}) it is difficult
to quantify this, or indeed draw any conclusions regarding the weaker
$\nu_1+2\nu_3$/$3\nu_3$ bands, without additional high-resolution experimental
data.

The largest source of error in our intensity calculations will undoubtedly be
the DMS. To improve on the CCSD(T)-F12b/cc-pVQZ-PP-F12 method by Hill
\etal~\cite{14HiPexx}, large CCSD(T)/aug-cc-pwCVnZ-DK ($n=4,5$) calculations
would likely be necessary (for an example, including additional post-CCSD(T)
corrections see Ref.~\onlinecite{13DePexx.AsH3}), which are currently computationally unmanageable
for a full 6D surface. Secondarily, the PES quality must be considered. Line
intensities are inexorably connected to the PES
through the wavefunctions in Eq.~(\ref{eq:intens}), and accurate modelling of
intensity transfer between lines (so-called `intensity stealing') relies on the
correct representation of rotation-vibration resonances in the PES.
Therefore, from a nuclear motion point-of-view, high-resolution measurements
complete with line intensities and quantum assignments, particularly for the
$800-1200$, $2900-3300$ and $5000-5300$ cm$^{-1}$ regions, would be most
beneficial for future modelling.

\section{Conclusion}\label{conclusion}
We have produced the first full-dimensional PES and DMS for the arsine molecule.
Both PES and DMS were computed at the CCSD(T)-F12b/cc-pVQZ-PP-F12 level of
theory, with implicit treatment of scalar relativistic effects via replacement
of 10 core electrons with a relativistic pseudopotential. A comparison with
standard CCSD(T)/aug-cc-pVQZ-DK based calculations employing the DKH8
Hamiltonian, showed that CCSD(T)-F12b/cc-pVQZ-PP-F12 level theory resulted in
significantly more accurate nuclear motion calculations.

Geometry optimisation and empirical adjustment of harmonic and certain cubic
terms in the pVQZ-PP-F12 PES resulted in $J=1-6$ rotational energies with a
root-mean-square error of 0.0055 cm$^{-1}$, and vibrational term values accurate
to within 1 cm$^{-1}$ for all reliably known experimental band centres under
6400 cm$^{-1}$. Utilising the empirical basis set correction scheme, 578
experimentally derived ($J=1-6$) rovibrational energies are reproduced with an
RMS of 0.122 cm$^{-1}$. Vibrational term value comparisons with eight
approximately known band centres showed that six agreed within 3.5 cm$^{-1}$
despite being omitted from the refinement. The remaining two displayed $\sim16$
cm$^{-1}$ discrepancies, most likely due to overlooked resonances.

Rotational-vibrational line intensity calculations were performed using the
refined PES and \text{ab initio} DMS, in conjunction with variational nuclear
motion calculations. The resulting line list, with full quantum assignments,
extends to 7000 cm$^{-1}$ and is complete up to 300 K. Comparisons with multiple
experimental sources show our intensity predictions to be reliable, in
particular, good overall agreement with the main absorption features present in
PNNL is noted. Our complete line list with quantum assignments is available from
the ExoMol website (www.exomol.com) in ExoMol format \cite{jt631}.

As far as we know, arsenic is the heaviest element for which there exists an
associated variationally-computed infrared molecular line list. Considering that
the quantum chemistry methods employed here are available for most p-block main
group elements~\cite{14HiPexx}, the outlook for studying similar systems in
future is positive.

\section*{Conflicts of interest}
There are no conflicts to declare.

\section*{Acknowledgements}
PAC thanks EPSRC for a CASE studentship under grant EP/L504889/1 and
Servomex for industrial sponsorship. The authors are grateful to Grant
J. Hill for his help with electronic structure calculations. SY and JT
thank STFC for support under grant ST/M001334/1.  The authors
acknowledge the use of the UCL Legion High Performance Computing
Facility (Legion@UCL), and associated support services, in the
completion of this work, along with the STFC DiRAC HPC Facility
supported by BIS National E-infrastructure capital grant ST/J005673/1
and STFC grants ST/H008586/1 and ST/K00333X/1..



\balance



\bibliographystyle{rsc} 


\providecommand*{\mcitethebibliography}{\thebibliography}
\csname @ifundefined\endcsname{endmcitethebibliography}
{\let\endmcitethebibliography\endthebibliography}{}

\end{document}